\documentclass[preprint,showpacs]{revtex4} 
\usepackage{natbib} 
\usepackage{epsfig}
\usepackage{dcolumn} 
\usepackage{bm}
\usepackage{amsmath}
\usepackage{amssymb}
\usepackage{multirow} 
\usepackage{float}

\begin{document}
\setcounter{MaxMatrixCols}{12}
\title{Hubbard-like Hamiltonian for ultracold atoms in a 1D optical lattice}
\author{Francesco Massel} \affiliation{Dipartimento di Fisica and UdR INFM,
  Torino Politecnico, Corso Duca degli Abruzzi 24, I-10129 Torino, Italy}
\author{Vittorio Penna} \affiliation{Dipartimento di Fisica and UdR INFM,
  Torino Politecnico, Corso Duca degli Abruzzi 24, I-10129 Torino, Italy}
\begin{abstract}
  Based on the standard many-fermion field theory, we construct models
  describing ultracold fermions in a 1D optical lattice by implementing a mode
  expansion of the fermionic field operator where modes, in addition to space
  localisation, take into account the quantum numbers inherent in local
  fermion interactions.  The resulting models are generalised Hubbard
  Hamiltonians whose interaction parameters are derived by a fully-analytical
  calculation.  The special interest for this derivation resides in its
  model-generating capability and in the flexibility of the trapping
  techniques that allow the tuning of the Hamiltonian interaction parameters
  over a wide range of values.  While the Hubbard Hamiltonian is recovered in
  a very low-density regime, in general, far more complicated Hamiltonians
  characterise high-density regimes, revealing a rich scenario for both the
  phenomenology of interacting trapped fermions and the experimental
  realization of devices for quantum information processing.  As a first
  example of the different situations that may arise beyond the models well
  known in the literature (the unpolarised-spin fermion model and the
  noninteracting spin-polarised fermion model), we derive a Rotational Hubbard
  Hamiltonian describing the local rotational activity of spin-polarised
  fermions. Based on a standard techniques we obtain the mean-field version of
  our model Hamiltonian and show how different dynamical algebras characterize
  the case of attractive and repulsive two-body potentials.
 \end{abstract}
 \maketitle
  \section{Introduction}
   \label{sec:intro}
   Since the Bose-Einstein condensation  of alkali atoms in
   magnetic traps \cite{Cornell, Ketterle}, a massive experimental and
   theoretical effort has been dedicated to the investigation of confined
   atoms in the extremely low-temperature regime (for a review see
   \cite{Leggett, Dalfovo, Jaksch, PS03}). 
   
   The flexibility of optical trapping techniques has suggested the devise of
   different configurations (lattices \cite{Anderson, Orzel,Morsch, Jaksch,
   Jaksch2}, superlattices, etc. \cite{Buonsante, Buonsante2, Peil, Guidoni, Roth,
   Santos}), opening a vast scenario of research.  The ability to tune atomic
   interactions via a magnetic field (Feshbach resonance \cite{Kokkelmans}),
   along with the proposal of single atom trap loading techniques
   \cite{CiracZoller}, has proven to be of capital importance for ultracold
   fermions physics, yielding the possibility to study fundamental aspects of
   superfluidity (BCS-BEC crossover, see e.g. \cite{Milstein, Giorgetti, Jin,
     Giorgini}) and envisaging new perspectives in quantum information
   processing \cite{Calarco,Zanardi}.
        
   The present work focuses on the theoretical investigation of the properties
   of (few) fermionic ultracold atoms loaded into a 1D optical lattice, where
   global confinement is ensured by a magnetic trap. The description of such a 
   physical system can be naturally performed in terms of a generalised
   Hubbard Hamiltonian (gHH) which is deduced from a general field-theoretic
   Hamiltonian with two body interaction \cite{GW98}. At this stage,
   particular care must be taken in the choice of the function basis for the
   field operator expansion. Although the symmetries of the system can provide
   selection rules that reduce the involvement of the gHH, the resulting
   coefficient structure is very rich and, as a direct consequence, the
   Hamiltonian hardly tractable. Nevertheless, the generality of the model
   gives rise to a wealth of sub-models, depending upon different
   approximations and regimes. The guideline to find simplified Hamiltonians
   is given by the thorough analysis of the gHH coefficient structure.
   
   From this perspective the analytical knowledge of the coefficients is a
   powerful tool to establish the physical relevance of different sub-models
   in the various situations that may be conceived in the framework of the
   trapped ultracold atoms physics.  Moreover, the nontrivial dependence of
   the coefficient from controllable external parameters provides the
   possibility to use these parameters to control the dynamics of the atoms
   trapped in the optical lattice. Thus the key aspect of this paper is the
   analytical determination of the hopping and interaction coefficients as a
   function of experimental parameters such as magnetic trap frequency, laser
   intensity, wavelength, angle between laser sources, $s$-wave scattering
   length etc.
  
   We would like to stress that the procedure followed here for the
   determination of the coefficients is statistics-independent: the bosonic or
   fermionic nature of the atoms loaded into the trap is completely taken into
   account by the commutators of raising and lowering operators that will be
   described in the paper. For example with the calculations performed here it
   seems feasible to go beyond the approximations that lead to the
   Bose-Hubbard model in the description of the BEC dynamics in optical
   lattices, taking into account the specific nature of the interaction
   between alkali atoms in a low density regime. Even for a ground-state
   calculation it can be shown that it is necessary to include levels beyond 
   the single particle ground state (see \cite{AB01}).  
      
   The confinement model considered here has a direct experimental
   relevance (see e.g. \cite{Modugno, Pezze}). However, while in \cite{Modugno,
   Pezze} a  number of atoms of the order of $10^4$ is considered,
   allowing thus the adoption of a semiclassical model, we focus on a low
   occupation-number regime similarly to what is done in \cite{Hofstetter} and
   \cite{Albus}, yet extending to a multi-band model whose correctness is
   limited by the validity limit of the space-mode approximation. 
   
   Challenging tasks for the future will include the determination of
   tractable yet interesting models for different aspects for theoretical
   condensed matter physics and quantum mechanics.  On the other hand the
   experimental realisation of systems that exhibit a behaviour which can be
   described in the framework of the various models here proposed, would
   represent an important achievement for both condensed matter
   experimentalist and theoreticians: the main difficulties seem to arise form
   the nearly-single atom trap loading and, quite naturally, from the coupling
   with the external environment.
   
   Throughout the paper we have tried to emphasise the generality of the
   procedure followed. However we have decided to write down and plot few
   numerical values of the coefficients to stress the fact that this
   calculation is a direct and relatively simple tool to shape out simplified
   and approximate Hamiltonians for different physical situations. 
   
   In section \ref{sec:FeTr} we depict the potential configuration of the
   system, moving then to the description of our field-theoretical approach.
   The field operators are written in terms of mode raising and lowering
   operators. Each mode corresponds to a set of quantum numbers, one of them
   identifies the lattice site (hence space-mode approximation) while the
   others describe on-site quantum numbers (local-mode)\cite{Illuminati}. As
   previously stated this choice is not unique, but symmetry constraints
   suggest expansions that emphasise conservation laws and selection rules.
   
   In section \ref{sec:HaCo} we evaluate the expression of the Hamiltonian 
   hopping and interaction coefficients and we try to describe the 
   interaction coefficient symmetry properties into some detail. 
   
   The purpose of section \ref{sec:SpCa} is twofold. One the one hand we show
   how, with suitable approximations, the Hamiltonian of the system reduces to
   known cases, such as the Hubbard Hamiltonian or a trivial non-interacting
   Hamiltonian. On the other we introduce a novel Rotational Hubbard
   Hamiltonian, as a first instance of the involvement of higher order
   approximations. For this case, by means of established mean-field
   approaches \cite{Lieb,Gilmore}, we suggest a possible path of research 
   involving general group-theoretical procedures \cite{Gilmore}. It will be 
   shown that these procedures, even if the explicit solution for the ground
   state is not given, allow to grasp interesting aspects of the physics of the 
   model here discussed.
  
   We have included two Appendices where the relatively simple but lengthy
   calculations of the tunnelling and interaction coefficients are provided
   explicitly. In Appendix \ref{sec:AppA} there are various plots of
   multilevel hopping parameters that supply a good example of the scenario
   that we are moving in and may constitute a good starting point for further
   investigation.

    \section{Fermions trapped in 1D optical lattices}
       \label{sec:FeTr}
    
     \subsection{General features}
      \label{sec:genFeat}

       The general field-theoretic Hamiltonian (see e.g. \cite{GW98}) with
       2-body interaction can be written as 
      \begin{equation}
        \label{eq:ft-hamiltonian}
        \hat{\mathsf{H}}=\int d\mathbf{r} \hat{\Psi}^\dagger(\mathbf{r})
        \mathsf{H}_{1b}(\mathbf{r})
        \hat{\Psi}(\mathbf{r}) 
        +\int d\mathbf{r} d\mathbf{r}' \hat{\Psi}^\dagger(\mathbf{r})
        \hat{\Psi}^\dagger(\mathbf{r}')
        \mathsf{H}_{2b}(\mathbf{r},\mathbf{r}') \hat{\Psi}(\mathbf{r}')
        \hat{\Psi}(\mathbf{r})
     \end{equation}
     where $\mathsf{H}_{1b}(\mathbf{r})$ represents the 1-body term of the
     Hamiltonian (kinetic + external potential term) while
     $\mathsf{H}_{2b}(\mathbf{r},\mathbf{r}')$ the 2-body interaction
     potential term, $\hat{\Psi}(\mathbf{r})$ is the field operator and
     $\hat{\Psi}^\dagger(\mathbf{r})$ its adjoint.
     
     As previously mentioned we will stick to neutral fermionic atoms loaded
     into a 1D optical lattice. The lattice is generated by two lasers
     counter-propagating along the $x$-axis, with wavenumber $K$.  The depth --
     or height, depending if red or blue detuning of the laser is considered --
     in each point $x$ the potential is proportional to the intensity of the
     laser and thus, according to the considered setup, to $\sin^2(2Kx)$, for
     the evaluation of the multiplicative constant see
     e.g. \cite{Leggett}. Here we set the multiplicative constant equal to
     $m\omega^2/(2K^2)$ where $\omega$ represents the harmonic oscillator
     frequency in the second order expansion of the term $V_{ext}$.  
  
     Global confinement is ensured by a cigar-shaped magnetic trap with
     principal axis along the $x$-direction (see e.g. \cite{Inguscio}). 
     This trap can be modelled by a 3D
     harmonic anisotropic trap of axial and radial frequencies equal to
     $\Omega_x$ and $\Omega_\perp$ respectively ($\Omega_x \ll \Omega_\perp$).
 
     The magneto-optical trap can be thought as if the constituents of the
     system were trapped in the cigar shaped potential with a ``slicing''
     effect of the laser, giving rise to a linear array of 3D prolate harmonic
     oscillators.  Besides, the radial trapping frequency has a deep influence
     on the interaction among the constituents of the system, allowing to
     control the volume of each ``disk''.
%%%%%%                
With the previous assumptions $\mathsf{H}_{1b}$ becomes
     $$
     \mathsf{H}_{1b}= E_{kin} + V_{ext}
     $$
     where 
     \begin{eqnarray}
        \label{eq:defV_ext}
          E_{kin} &=& -\frac{\hbar^2 \nabla^2}{2m}  \nonumber   \\
          V_{ext} &=&  \frac{m}{2}  
                        \left[ 
                              \Omega_x^2 x^2 + \Omega_\perp^2 \rho^2
                        \right]  + \frac{m \omega^2}{2K^2} \sin^2(Kx)\, ,
      \end{eqnarray}
      and the second term of $V_{ext}$ represents the harmonic
      confinement of the magnetic trap, while the third one corresponds to the
      optical potential and $\rho^2=y^2+z^2$.
%%%%%%%      
      For future convenience, we write equation \eqref{eq:defV_ext} as 
      \begin{equation}
        \label{eq:H_1body}
          \mathsf{H}_{1b}=E_{kin}+\sum_jV_j+\left(V_{ext}-\sum_jV_j\right) 
      \end{equation}
      with
      \begin{equation}
        \label{eq:V_j}
        V_j=\Pi_j(x) \frac{m}{2} \left[ \Omega_x^2 j^2\frac{\pi^2}{k^2}+ 
            \omega^2 x_j^2+
         \Omega_\perp^2 \rho^2 \right]
         \end{equation}
        $\Pi_j(x)=\Pi\left(\frac{Kx}{\pi}-j\right) $ where   $\Pi(x)$
         is the rectangle function ($\Pi(x)=1$ for $-1 \leq x < 1$,
         $\Pi(x)=0$ elsewhere), $k=l_\perp K$ (with $l_\perp=\sqrt{\hbar/(m
         \omega_\perp)})$ and  
         $x_j=\left(x-j \frac{\pi}{k}\right)$.  
         Here the harmonic axial 
         confinement of the magnetic field has been considered as a
         site-dependent -- with $j$ site index -- constant addictive term,
         merely shifting the local minima of the optical potential.
%%%%%%%%%%%%%
         From Eq. \eqref{eq:H_1body} with the properties of the rectangle
         function we obtain
  \begin{equation}
        \label{eq:H_1body2}
          \mathsf{H}_{1b}= \sum_j \Pi_j(x)
                            \left[
                                  \left(E_{kin}+V_j\right)+
                                  \left(V_{ext}-V_j\right) 
                            \right]
  \end{equation}
%%%%%%%
  Hereafter the axial confinement of the magnetic trap will be neglected
  (small $\Omega_x$).
  
  We are now led to consider two different terms in Eq. \eqref{eq:H_1body2}.
  The first represents a local harmonic-oscillator Hamiltonian
  \begin{eqnarray}
    \label{eq:Harm_Ham}
    \mathsf{H}_{j}^{ho}&=& \Pi_j(x)
    \left(E_{kin}+V_j\right)    \nonumber \\
    &=& \Pi_j(x) \left[ \frac{\hbar^2 \nabla^2}{2m}+
      \frac{m\omega^2}{2} x_j^2+ \frac{m \Omega_\perp^2}{2}\rho^2 \right]
  \end{eqnarray}
  and a hopping one
  \begin{eqnarray}
    \label{eq:Tunn_Ham}
     \mathsf{H}_j^{tunn} &=&  \Pi_j(x)
                             \left(V_{ext}-V_j\right) \nonumber \\
                         &=&  \Pi_j(x) 
                            \left[ 
                                  \frac{m \omega^2}{2K^2} \sin^2(Kx)-
                                  \frac{m\omega^2}{2} 
                                  x_j^2
                            \right].
  \end{eqnarray}
%%%%%%  
  The term $V_j$ is the local second-order expansion of the optical potential,
  thus equation \eqref{eq:Tunn_Ham} represents the discrepancy between an
  harmonic potential and the true optical potential, describing hopping of
  atoms between neighbouring sites.
  
  Neutrality of the atoms, ensuring a finite-range interaction allows us to
  introduce a pseudo-potential approximation (see e.g. \cite{KH87})
  \begin{equation}
    \label{eq:PseudoPot}
     U(\mathbf{r})=  \sum_{j}
                     \Pi_j(x) 
                     {\tilde a}_s
                     \delta(\mathbf{r})
                     \frac{\partial}{\partial \mathbf{r}}\mathbf{r}\, ,
  \qquad {\tilde a}_s := \frac{4\pi \hbar^2 a_s}{m}\, , 
  \end{equation}
  where $\mathbf{r}$ is the interatomic distance and $a_s$ the $s-$wave
  scattering length ($a_s$ in our approximation is considered constant).  The
  validity of this model is ensured by the low energies involved in these
  interactions, direct consequence of both low temperature limit (virtually
  zero) and diluteness (low Fermi energy). Besides, the form of Eq.
  \eqref{eq:PseudoPot} shows that on-site terms only will contribute to the
  interaction Hamiltonian.
  Thus Eq. \eqref{eq:ft-hamiltonian} can be rewritten in the form
  \begin{multline}
    \label{eq:ft-hamiltonian2}
     \hat{\mathsf{H}}= 
%\\ 
                                        \sum_j  \Bigl [
                                         \int d\mathbf{r} 
                                            \hat{\Psi}^\dagger(\mathbf{r})
                                       \Bigl (  \mathsf{H}_j^{ho}(\mathbf{r})
                                    +\mathsf{H}_j^{tunn}(\mathbf{r}) \Bigr )
                                            \hat{\Psi}(\mathbf{r}) +  \\ 
%                                     + \int d\mathbf{r} 
%                                           \hat{\Psi}^\dagger(\mathbf{r})
%                                             \mathsf{H}_j^{tunn}(\mathbf{r})
%                                           \hat{\Psi}(\mathbf{r}) + \right. 
                                   \! + {\tilde a}_s
                                              \int d\mathbf{r} d\mathbf{r}' 
                                                \hat{\Psi}^\dagger(\mathbf{r}) 
                                           \hat{\Psi}^\dagger(\mathbf{r}')        
                                                    \delta(\mathbf{r}-\mathbf{r}')
% \left.      
                                \hat{\Psi}(\mathbf{r}') \hat{\Psi}(\mathbf{r})
                              \Bigr ].  
\end{multline}

\subsection{The (space+local)-modes expansion}

  The choice of the basis for the expansion of the field operators is crucial.
  As already suggested by the grouping of terms in Eq. \eqref{eq:Harm_Ham}, we
  will choose a basis constituted by local harmonic oscillator eigenfunctions.
    In addition, because of the symmetry of the system we have chosen
  central-symmetric 2D h.o. eigenfunctions for the 2D isotropic radial
  h.o.\cite{Wunsche}, instead of decomposing it in 1D h.o. eigenfunctions,
  this will give us deeper insight into conservation laws and selection rules
  imposed by the symmetries of the system.  
  We then have
  \begin{equation}
    \label{eq:SpModeExp}
      \hat{\Psi}(\mathbf{x})=
          \sum_{i,n_x,J,m,\sigma} 
             u_{n_x}(x-x_i)\mathcal{L}_{J,m}(\rho,\phi)\xi(\sigma)
           \hat{c}_{n_x,J,m,i,\sigma}     
\end{equation}
with
\begin{eqnarray}
    \label{eq:HO}
     u_{n}(x) &=& \frac{1}{\sqrt{2^n n! \sqrt{\pi}l_x}} 
                   H_n(x/l_x)e^{-\frac{x^2}{2l_x^2}} \, ,   \\ 
     \mathcal{L}_{J,m}(\rho,\phi) &=& \frac{e^{2 i m \phi}}{\sqrt{\pi}l_{\perp}}
                            \, C_{Jm}
                            \left(
                                  \frac{\rho}{l_\perp} \right)^{2m} 
                                  L_{J-m}^{2J} \left( {\rho}/{l_\perp} \right)
\, , \,\,\,
\end{eqnarray}
$C_{Jm}= \sqrt{{(J+m)!}/{(J-m)!}}$, $\xi(\sigma)$ is a spin function and
$l_x=\sqrt{\hbar/(m\omega_x)}$.
In this decomposition $u_{n}(x)$ is a 1D harmonic-oscillator eigenfunction
($H_n$ represent the $n$th Hermite polynomial) and $\mathcal{L}_{J,m}$ a
2D harmonic-oscillator eigenfunction \cite{Wunsche} with $L_{J-m}^{2J}(x)$ a generalised Laguerre polynomial.  
  
  Fermionic operators will thus have 5 indexes: 3 of them ($n_x,J,m$) identify
  (2+1)D local harmonic oscillator states, while $i$ identifies the site and
  $\sigma$ the spin.
  While $n_x$ has its usual interpretation of 1D harmonic oscillator number
  operator eigenvalue, $J$ and $m$ can be construed as angular momentum and
  $x-$axis component of the angular momentum, respectively.
  
  This decomposition can be thought as a generalised space-mode approximation, with
  additional local modes that, in the present case, correspond to the local (2+1)D
  harmonic-oscillators quantum numbers. If not explicitly required, we will
  use $\phi_\alpha= u_{n_\alpha}(x-x_{i_\alpha})
  \mathcal{L}_{J_\alpha, m_\alpha}(\rho,\phi)$
  $\xi(\sigma_\alpha)$, with
  $\alpha=\{n_\alpha,J_\alpha,m_\alpha,i_\alpha,\sigma_\alpha\}$ to simplify
  the index notation.
  We wish to stress that decomposition \eqref{eq:SpModeExp} is an
  \textit{approximation} of field $\hat{\Psi}(\mathbf{x})$: there is a 
  non-nil overlapping between wavefunctions
  belonging to different sites, thus orthogonality is not fulfilled.
  Nevertheless these overlapping integrals are supposed to be small, ensuring the
  consistency of this choice \cite{Jaksch2}.
  
  In the forthcoming calculation of the interaction
  term, Eq. \eqref{eq:HO} allows us 
  to easily recognise that $m$ is a conserved quantity.  If we come back
  to \eqref{eq:ft-hamiltonian}, with the decomposition \eqref{eq:SpModeExp} we
  obtain
% \begin{multline}
%   \label{eq:ft-hamiltonian3}
%   \hat{\mathsf{H}}=\sum_{j}\left[ \sum_{\alpha,\beta} \int d\mathbf{r}
%     \phi_\alpha^*(\mathbf{r}) \mathsf{H}_j^{h.o.}  \phi_\beta(\mathbf{r})
%     \hat{c}^\dagger_\alpha \hat{c}_\beta+   \right.\\
%   \sum_{\alpha,\beta} \int d\mathbf{r} \phi_\alpha^*(\mathbf{r})
%   \mathsf{H}_j^{tunn} \phi_\beta(\mathbf{r})
%   \hat{c}^\dagger_\alpha \hat{c}_\beta+  \\
%    \frac{4\pi \hbar^2a_s}{m}  \sum_{\alpha,\beta,\gamma,\delta} \int
%   d\mathbf{r} d\mathbf{r}' \phi_\alpha^*(\mathbf{r})
%   \phi_\gamma^*(\mathbf{r}') \delta(\mathbf{r}-\mathbf{r}')
%   \phi_\beta(\mathbf{r}) \phi_\delta(\mathbf{r}') \cdot\\
% \left. \cdot \hat{c}^\dagger_\alpha \hat{c}^\dagger_\beta \hat{c}_\delta 
% \hat{c}_\gamma \right]
%  \end{multline}
%
\begin{eqnarray}
    \label{eq:ft-hamiltonian3}
    \hat{\mathsf{H}}=\sum_{j} \sum_{\alpha,\beta}
    \left[ \,  \int d\mathbf{r} \,
      \phi_\alpha^*(\mathbf{r}) \mathsf{H}_j^{ho}  \phi_\beta(\mathbf{r})
      \hat{c}^\dagger_\alpha \hat{c}_\beta +   
 \int d\mathbf{r} \, \phi_\alpha^*(\mathbf{r})
    \mathsf{H}_j^{tunn} \phi_\beta(\mathbf{r})
    \hat{c}^\dagger_\alpha \hat{c}_\beta + \right. \nonumber \\ 
   \left. {\tilde a}_s \sum_{\gamma,\delta} \int
    d\mathbf{r} d\mathbf{r}' \, 
   \phi_\alpha^*(\mathbf{r}) \phi_\gamma^*(\mathbf{r}') 
   \times  \delta(\mathbf{r}-\mathbf{r}')
    \phi_\beta(\mathbf{r}) \phi_\delta(\mathbf{r}') 
\, \hat{c}^\dagger_\alpha \hat{c}^\dagger_\beta \hat{c}_\delta
      \hat{c}_\gamma \right]
  \end{eqnarray}

%%%%%%%%%%%%%%%%%%%%%%%%%%%%%%%%%%%%%%%%%%%%%%%%%%%%%%%%%%%%%%%%%%%%%%%%%%%%%%%%
\section{Hamiltonian Coefficients}
      \label{sec:HaCo}
We are now in the position to calculate all the coefficients in
Hamiltonian \eqref{eq:ft-hamiltonian3}.  The first term becomes
\begin{equation}
      \label{eq:lambda}
       \hat{\mathsf{H}}^{ho}=\sum_{j,\alpha,\beta} 
             \lambda_\beta  
               \int_{-\infty}^{\infty} d\mathbf{r}
                   \, \Pi_j(x)
                     \phi_\alpha^*(\mathbf{r})
                     \phi_{\beta}(\mathbf{r}) 
             \hat{c}^\dagger_\alpha\hat{c}_\beta \, , 
\end{equation}
    where $\lambda_\beta$ is the (2+1)D harmonic-oscillator eigenvalue
    \begin{equation}
      \label{eq:lambda2}
       \lambda_{n_\beta,J_\beta,m_\beta,i_\beta,\sigma_\beta}= 
          \left[
                \hbar\omega_x
              \left(
                  n_\beta+\frac{1}{2}
              \right)+  
                \hbar\left(2J_\beta+1\right)
          \right] \, .
    \end{equation}
Eq. \eqref{eq:lambda} can be written as
    \begin{equation}
      \label{eq: lambda3}
      \sum_{j,\alpha,\beta}
           \lambda_\beta 
           \delta_{\alpha,\beta} \delta_{i_\alpha,j} 
           \hat{c}^\dagger_\alpha\hat{c}_\beta =
      \sum_{\alpha} \lambda_\alpha \hat{n}_\alpha   
    \end{equation}
where the second Kronecker delta is a consequence of the space-mode
    approximation, i.e. we consider only superposition of wavefunctions among
    which at least one is a local harmonic-oscillator eigenfunction, while the
    first one stems from the orthogonality of the $\phi_\gamma(x)$ functions.
    
\begin{figure}[htbp]
      \centering \epsfig{figure=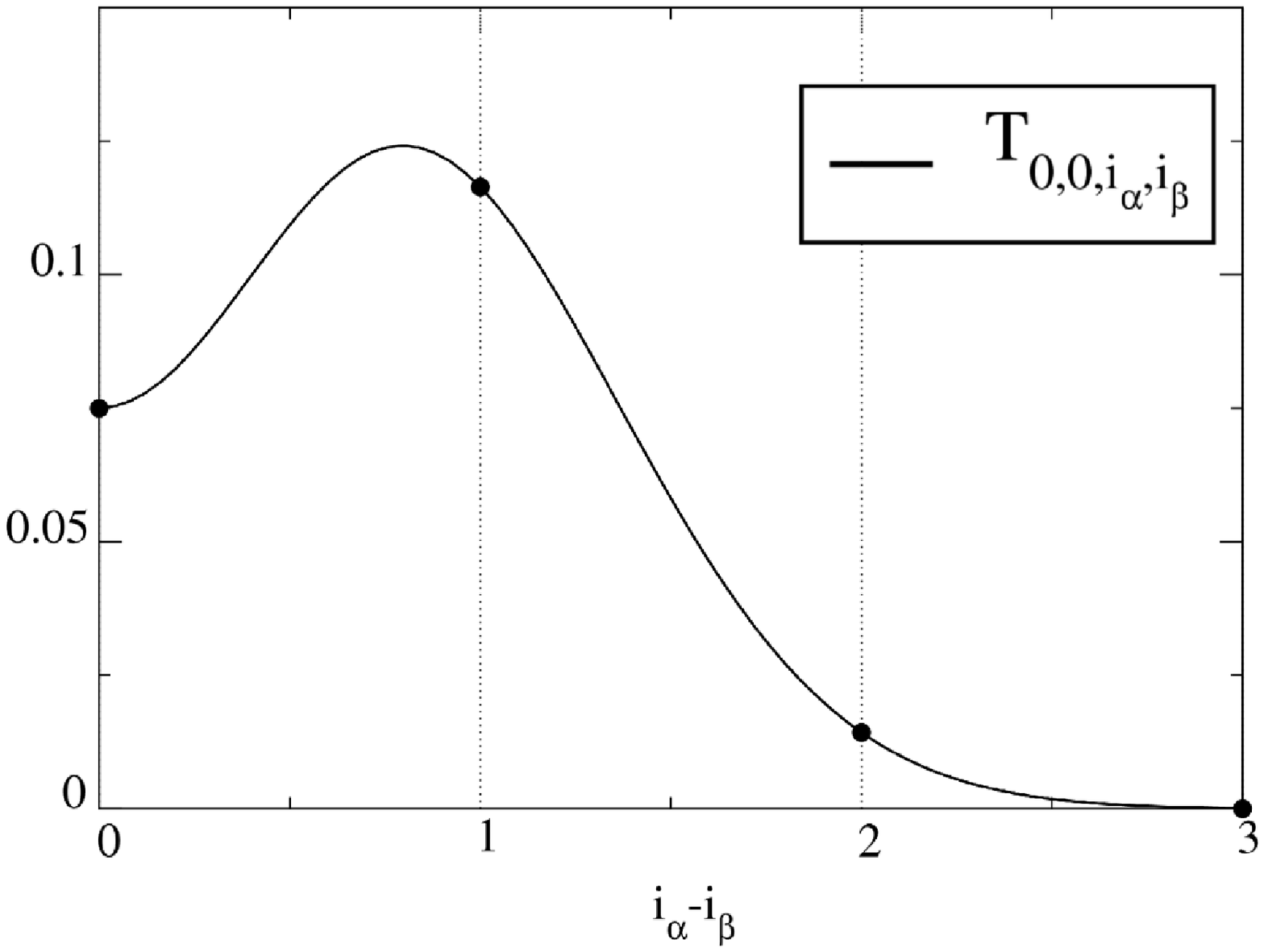,angle=270,width=0.4\textwidth}
      \centering \epsfig{figure=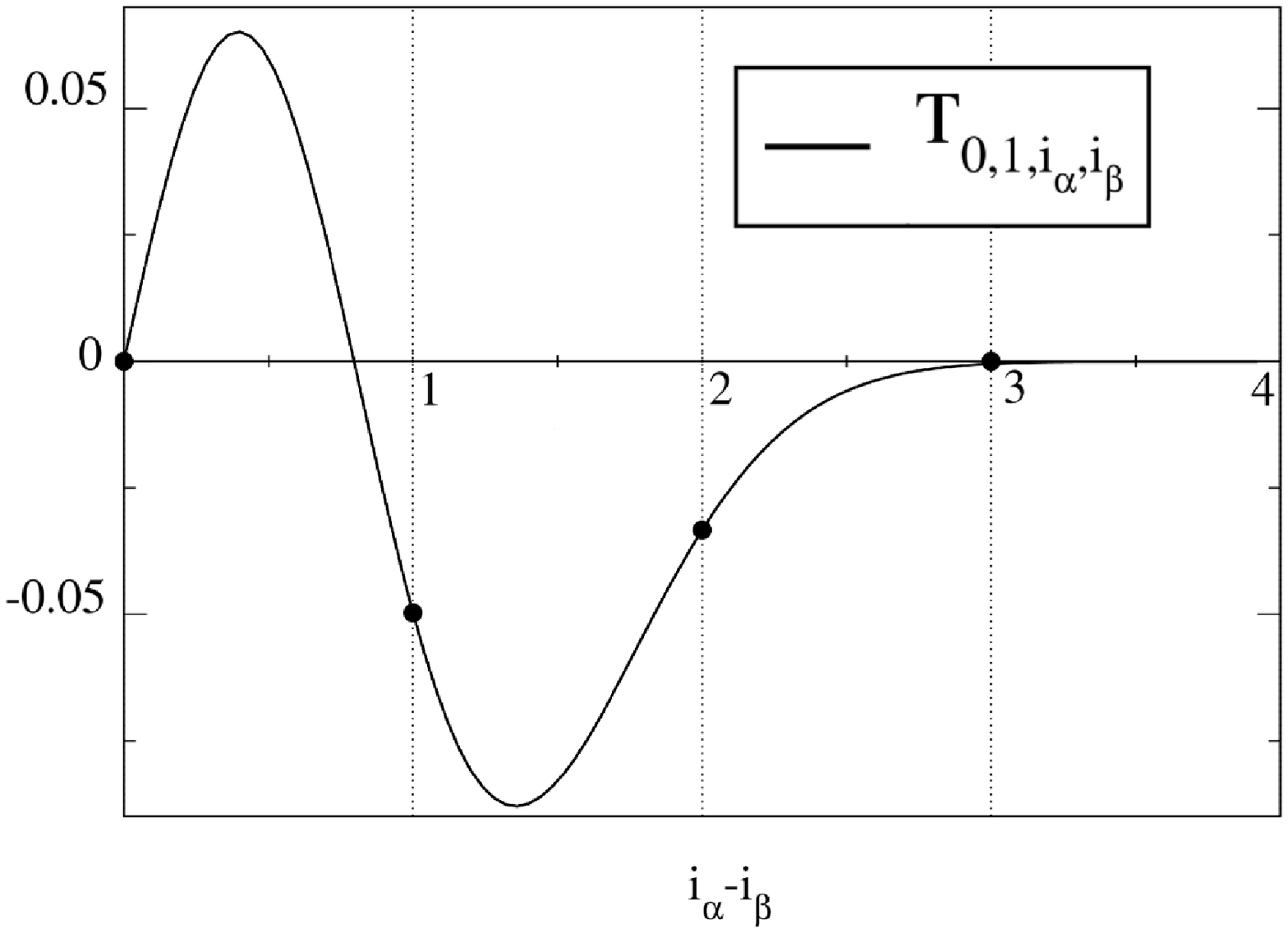,angle=270,width=0.4\textwidth}
      \centering \epsfig{figure=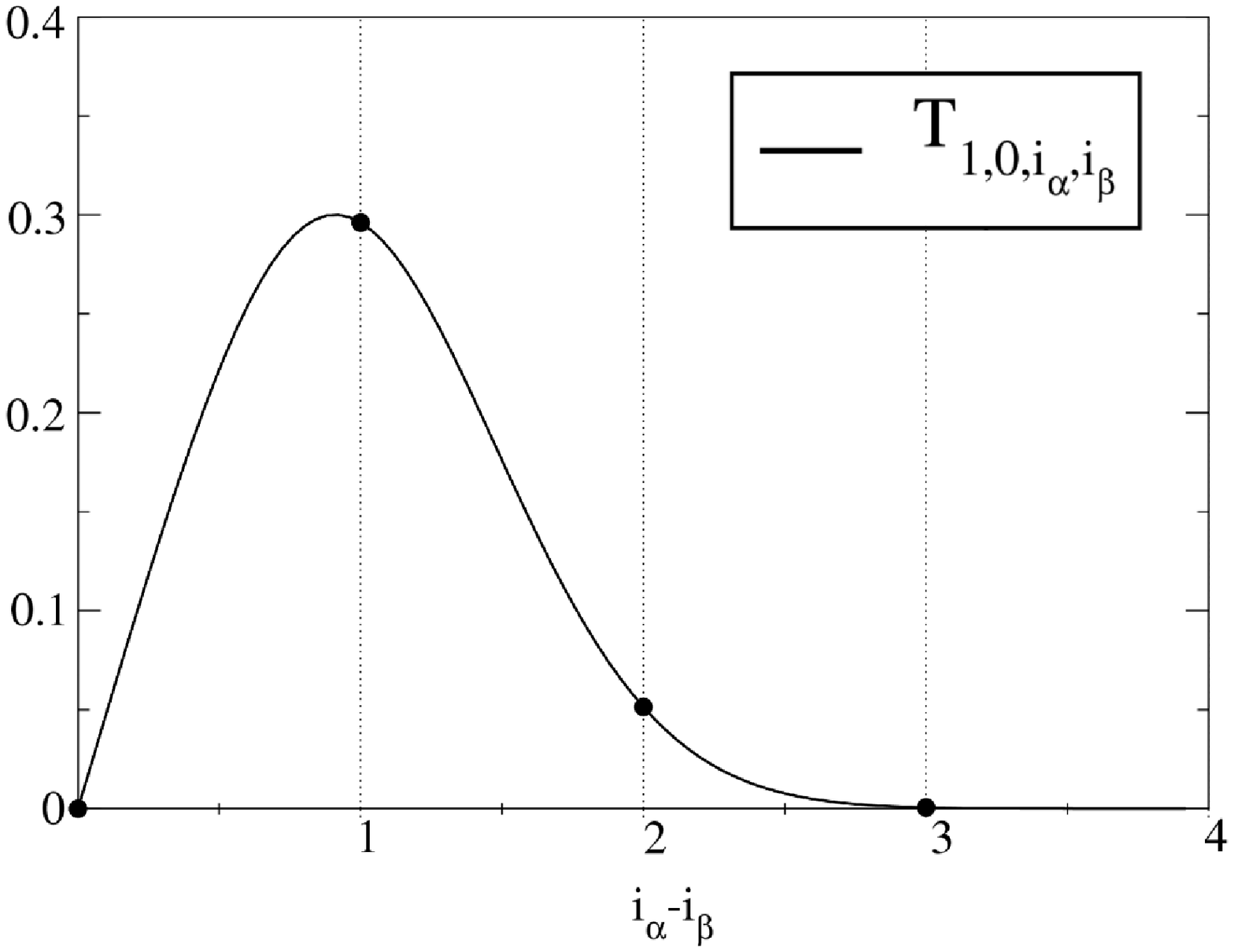,angle=270,width=0.4\textwidth}
      \centering \epsfig{figure=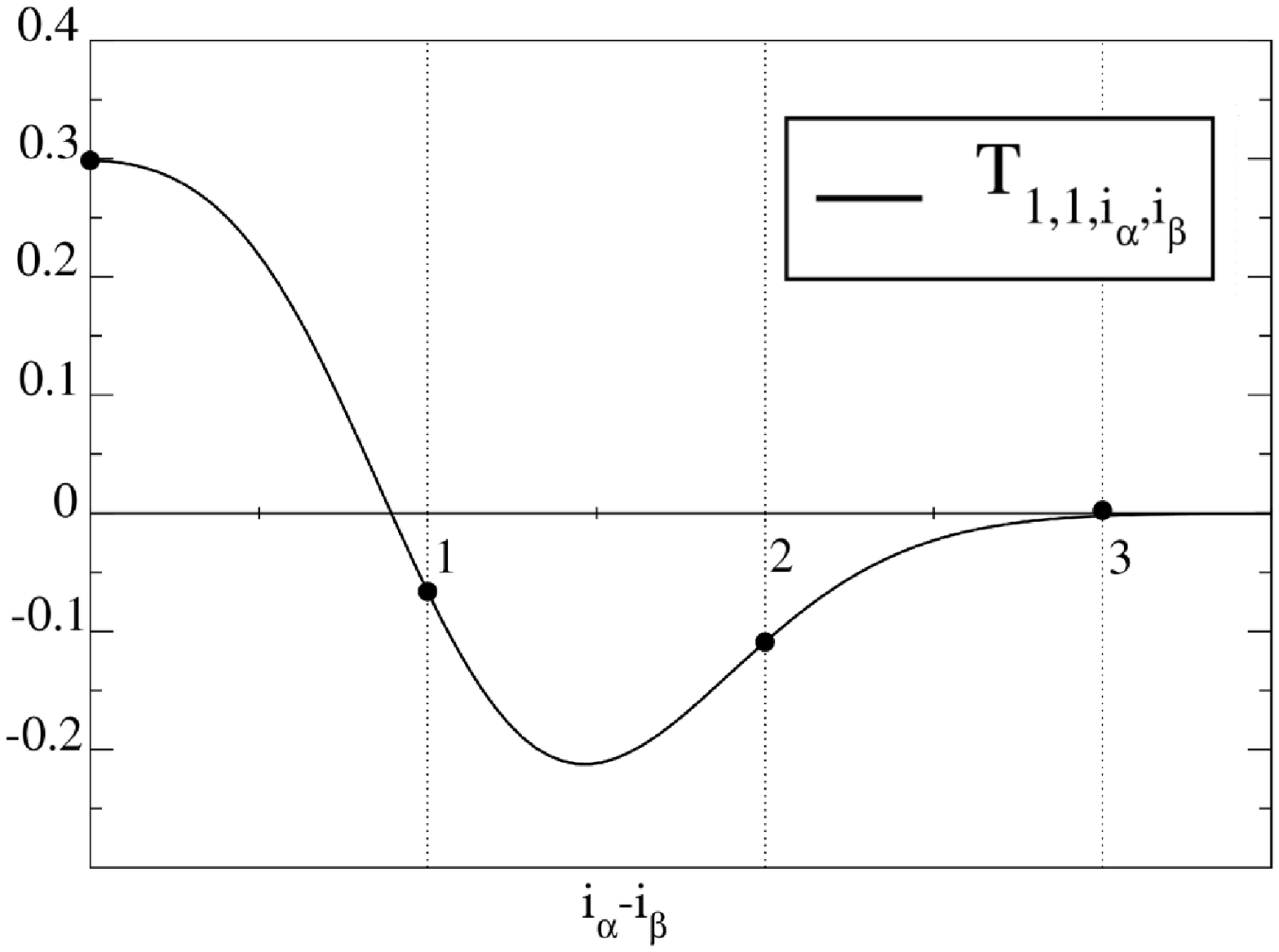,angle=270,width=0.4\textwidth} 
      \caption{Plots of the tunnelling coefficients from $T_{00}$ to $T_{22}$.
      A detailed discussion of the analytic expression of the hopping
      parameter will be given in Appendix \ref{sec:AppA}.}
        \label{t00}
    \end{figure}

    We move now to the evaluation of the integral in the second term of
    equation \eqref{eq:ft-hamiltonian3}. Namely
\begin{equation}
      \label{eq:T}
       \hat{\mathsf{H}}^{tunn}= \sum_{j,\alpha,\beta} 
          \int d\mathbf{r} \,
              \phi_\alpha^*(\mathbf{r})
                    \Pi_j(x) 
                    \mathsf{H}_j^{tunn}(x) 
              \phi_\beta(\mathbf{r})
\end{equation}
being $\mathsf{H}_j^{tunn}(x)$ independent of radial and spin degrees of
    freedom, we can rewrite equation \eqref{eq:T} as
    \begin{equation}
      \label{eq:T2}
      \hat{\mathsf{H}}^{tunn}= \sum_{j,\alpha,\beta}
                                   \delta_{J_\alpha,J_\beta} 
                                   \delta_{m_\alpha,m_\beta}
                                   \delta_{\sigma_\alpha,\sigma_\beta} 
                                   \delta_{i_\alpha,j}
                                   \int dx \,  u^*_{n_\alpha,i_\alpha}(x) 
                                                  \mathsf{H}_j^{tunn}(x)
                                          u_{n_\beta,i_\beta}(x) 
                                  \hat{c}^\dagger_\alpha\hat{c}_\beta.
    \end{equation}
    With the same assumptions of the local harmonic-oscillator case the
    integral in equation \eqref{eq:T2} becomes
\begin{equation}
      \label{eq:T3}
   K_{n_{\alpha}, n_{\beta}} \hbar \omega_x     \int dy \,
   e^{-\frac{(y-\tau)^2}{2}}  H_{n_\alpha}\left(y-\tau\right) 
                        \mathsf{H}_j^{tunn}(y) 
  e^{-\frac{y^2}{2}}  H_{n_\beta} \left ( y \right)\, ,
\end{equation} 
where $K_{n_{\alpha}, n_{\beta}} = 
[ 2^{n_\alpha+n_\beta } n_\alpha!\, n_\beta!\, \pi ]^{-1/2}$ 
and we have put $y=(x-i_\beta d)/l_x$ (where $d=\pi/k$), $\tau=(i_\beta-i_\alpha)$
and $\Omega=Kl_x$.  
By substituting the expression of $\mathsf{H}_j^{tunn}(y)$
from equation Eq. \eqref{eq:Tunn_Ham} 
we obtain
\begin{equation}
      \label{eq:T4}
  K_{n_{\alpha}, n_{\beta}} \hbar \omega_x
%\frac{\hbar \omega_x} {\sqrt{2^{n_\alpha+n_\beta}n_\alpha!n_\beta!\pi}}
        \int dy \, e^{-\frac{(y-\tau)^2 +y^2}{2}}
                  H_{n_\alpha}\left(y-\tau\right) H_{n_\beta}\left(y\right) \\ 
                           \left[ 
                                  \frac{1-\cos(2\Omega y)}{4\Omega^2} 
                                  - \frac{y^2}{2} \right] 
       \hat{c}^\dagger_\alpha\hat{c}_\beta.
\end{equation}
If we define
\begin{multline}
    \label{eq:T5}
    T_{\alpha,\beta} = \frac{ K_{n_{\alpha}, n_{\beta} } }{2}
\delta_{J_\alpha,J_\beta} \delta_{m_\alpha,m_\beta} 
\delta_{ \sigma_\alpha, \sigma_\beta} \hbar \omega_x
     \int dy \, e^{-\frac{ (y-\tau)^2 + y^2 }{2} } \quad \\
\times     H_{n_\beta}\left(y\right)  H_{n_\alpha}\left(y-\tau\right)
     \left[\frac{y^2}{2} - \frac{1-\cos(2\Omega y)} {4\Omega^2}  \right]
\end{multline}
equation \eqref{eq:lambda} 
becomes
\begin{equation}
      \label{eq:T6}
        \hat{\mathsf{H}}^{tunn}= 
                                 -\sum_{\alpha,\beta} 
                                     T_{\alpha,\beta}
                                  \hat{c}^\dagger_\alpha\hat{c}_\beta \, .
\end{equation}
Then the term 
$T_{\alpha,\alpha}\hat{c}^\dagger_\alpha\hat{c}_\alpha$ can be
incorporated into the $\hat{\mathsf{H}}^{h.o.}$ term, giving
    \begin{equation}
      \label{eq:mu}
       \mu_\alpha= \lambda_\alpha - T_{\alpha,\alpha}.
    \end{equation}
We will here skip the explicit solution of the integral in Eq.
    \eqref{eq:T5}, along with the analytic expression of T, which can be found
    in Appendix \ref{sec:AppA}.  These calculations allow us to write
    \begin{equation}
      \label{eq:Tdelta}
      T_{\alpha,\beta}=\delta_{J_\alpha,J_\beta}
                       \delta_{m_\alpha,m_\beta} 
                       \delta_{\sigma_\alpha,\sigma_\beta}
                       T_{n_\alpha,n_\beta,i_\alpha,i_\beta}.
    \end{equation}
    In fig. \ref{t00} we have the plot the coefficient
    $T_{n_\alpha,n_\beta,i_\alpha,i_\beta}$ as a function of the ratio between
    distance and the period of the optical lattice, for
    $n_\alpha,n_\beta=0,1$. In boldface we have marked the points
    corresponding to discrete values of the ratio $x/d$, i.e. the points with
    a relevant physical meaning, The values of $T$ plotted here are in
    arbitrary units. Even if the correctness of the above procedure seems
    undoubted, it must be remembered that it is entirely based on the
    space-mode approximation, whose validity depends on the overlapping of
    wavefunctions belonging to different sites and thus might be violated.
    
    These plots show how the tunnelling amplitude varies with the distance. In
    particular it is clear how, for long-distance tunnelling, there is a
    negative exponential dependence. Nevertheless, if the experimental
    conditions are properly chosen (i.e. angle between counterpropagating
    laser beams and their power), it is possible to obtain conditions where, for
    instance nearest-neighbour and next-to-nearest neighbour tunnelling
    coefficients have opposite signs (see e.g.  Fig.\ref{t0xf}
    $T_{0,0,i_\alpha,i_\beta}$), and thus the model, in that case, might
    exhibit frustration.

     We will now move to the determination of the interaction term, namely the
     last term of equation \eqref{eq:ft-hamiltonian3}.
As a first step, we can write the integral in cylindrical coordinates
\begin{eqnarray}
      \label{eq:U1}
%\frac{4\pi \hbar^2a_s}{m}
              &&{\tilde a}_s   \int d\mathbf{r}d\mathbf{r}' 
                             \phi_\alpha^*(\mathbf{r})
                             \phi_\gamma^*(\mathbf{r}') 
                                    \delta(\mathbf{r}-\mathbf{r}') 
                             \phi_\beta(\mathbf{r}) 
                             \phi_\delta(\mathbf{r}')          =   \nonumber\\
%\frac{4\pi \hbar^2a_s}{m}   
              && {\tilde a}_s  \int dx dx'  
                                 u^*_{n_\alpha}(x-x_{i_\alpha})
                                   u^*_{n_\gamma}(x'-x_{i_\gamma}) 
                            \delta(x-x') 
                                 u_{n_\beta}(x-x_{i_\beta}) 
                                 u_{n_\delta}(x'-x_{i_\delta})  \nonumber \\
              && \times  \int d\tilde{\rho} d\tilde{\rho}' \, 
                           \frac{\tilde{\rho}}{\pi}  
                              \int d\phi d\phi' \,
                     \mathcal{L}^*_{J_\alpha,m_\alpha}(\tilde{\rho},\phi )
                                 \mathcal{L}^*_{J_\gamma,m_\gamma}(\tilde{\rho}',\phi')
                                \delta(\rho-\rho')  \nonumber \\
                             && \delta(\phi-\phi') 
                        \mathcal{L}_{J_\beta,m_\beta}(\tilde{\rho},\phi)   
                       \mathcal{L}_{J_\delta,m_\delta}(\tilde{\rho}',\phi')         
\end{eqnarray}
    with $\tilde{\rho}=\rho/l_\rho$ and the identity 
    $$
      \delta(\mathbf{r})=\frac{\delta(\rho)\delta(\phi)}{\pi \rho}.
    $$

    As we are dealing with a short range interaction modelled by a
    $\delta(\mathbf{r}-\mathbf{r'})$ function, we will consider on-site
    interaction only
    ($\tilde{x}_{i_\alpha}=\tilde{x}_{i_\beta}=\tilde{x}_{i_\delta}=\tilde{x}_{i_\gamma}$).
    
    This choice is completely justified because the interaction term is
    modelled by a pseudopotential term for which nearest-neighbours interactions
    become negligible.  
    In this case the first integral on the left-hand
    side of Eq. \eqref{eq:U1} becomes
\begin{eqnarray}
     \label{eq:U2}
     U_x=
\frac{1}{\pi l_{x}}\sqrt{\frac{2^{-(n_\alpha+n_\beta+n_\gamma+n_\delta)}}
              {n_\alpha!n_\beta!n_\gamma!n_\delta!}}
%%K_{n_\alpha, n_\beta} K_{n_\gamma, n_\gamma}
             \int d\tilde{x} \,
                H_{n_\alpha}(\tilde{x})
                H_{n_\beta}(\tilde{x}) 
                H_{n_\gamma}(\tilde{x})
                H_{n_\delta}(\tilde{x})
                e^{-2\tilde{x}^2}
\end{eqnarray}
with $\tilde{x}=x/l_x$, whose explicit calculation is given in 
Appendix \ref{sec:AppB}. Here we just give the final result
\begin{equation}
   \label{eq:U3}
   U_x= \frac{\delta_{\|\bar{n}\|,2 \mathbb{N}}}
         {\pi l_x}
            \sum_{\bar{s}}^{\bar{n}} 
                \frac{\Xi ({\bar{s}} ) }{\sqrt{2}^{\|\bar{s}\|+3}} 
                      \, \Gamma  \left[
                                   \frac{\left(\|\bar{n}\|-\|\bar{s}\|\right)}{2}+1
                              \right]
\end{equation}
with
\begin{equation}
      \label{eq:Gamma}
        \Xi\left(\bar{s}\right) = \left\{
            \begin{array}{ll}
              0  & \textrm{if $s_\theta$ odd}  \\
                   \prod_\theta \frac{ 1}{n_\theta!} {n_\theta \choose s_\theta} 
                           H_{s_\theta}(0) 
                 & \textrm{if $s_\theta$ even} \\
            \end{array}
        \right. 
\end{equation}
The summation is to be intended as 4 separate summations over
    the components of a vector
    $\bar{s}=\left\{s_\alpha,s_\beta,s_\gamma,s_\delta\right\}$ from
    $\left\{0,0,0,0\right\}$ to
    $\bar{n}=\left\{n_\alpha,n_\beta,n_\gamma,n_\delta\right\}$.  The norm
    $\|\bar{x}\|$ is a 1-norm ($\|\bar{x}\|=\sum_\theta |x_\theta|$,
    $\theta=\alpha,\beta,\gamma,\delta$) and the $\delta$ in Eq. \eqref{eq:U2}
    represents the parity selection rule, obtained from the explicit
    calculation of the integral.
    
For the radial part of the integral we have
\begin{equation}
  \label{eq:U5}
  U_\rho=\int\int d\rho d\phi \, \,
               \frac{\rho}{\pi} \mathcal{L}^*_{J_\alpha,m_\alpha}(\rho,\phi)
                       \,  \mathcal{L}^*_{J_\gamma,m_\gamma}(\rho,\phi) 
                          \mathcal{L}_{J_\beta,m_\beta}(\rho,\phi)
                       \,  \mathcal{L}_{J_\delta,m_\delta}(\rho,\phi)
\end{equation}
with the definition given by Eq. \eqref{eq:HO}, we can easily perform the
 angular integration and we obtain
 \begin{equation}
  \label{eq:U6}
   U_\rho=\frac{2\delta_{m_\alpha+m_\gamma,m_\beta+m_\delta}}
               {\pi^2} 
          \int_0^\infty \! d\rho \rho \,
                R^*_{J_\alpha,m_\alpha}(\rho)
                \, R^*_{J_\gamma,m_\gamma}(\rho) 
                \, R_{J_\beta,m_\beta}(\rho)
                \, R_{J_\delta,m_\delta}(\rho).
\end{equation}
The reader is again addressed to Appendix \ref{sec:AppB} 
for the explicit evaluation of the integral in Eq. \eqref{eq:U6}.  
The result is given by
\begin{equation}
  \label{eq:U7}
  U_\rho= \frac{\delta_{m_\alpha+m_\gamma,m_\beta+m_\delta}}
               {\pi^2 l_\perp^2} 
           \sum_{\bar{q}=\bar{|m|}}^{\bar{J}}
             \Lambda\left(\bar{J},\bar{m},\bar{q}\right) 
              \frac{\Gamma \left(\|\bar{q}\|+3/2\right)}
                   {2^{\|\bar{q}\|+3/2}}  
\end{equation}
with
\begin{equation}
  \label{eq:Lambda}
   \Lambda\left(\bar{J},\bar{m},\bar{q}\right)=
             \prod_{\theta=\alpha,\beta,\gamma,\delta}
                \frac{(-1)^{J_\theta-q_\theta}
                           \sqrt{(J_\theta+m_\theta)!
                            (J_\theta-m_\theta)!}}
                     {(J_\theta-q_\theta)!
                      (q_\theta+m_\theta)!
                      \left(q_\theta-m_\theta\right)!}
\end{equation}
and, following previous notation, we obtain 
$\bar{q}=\left\{q_\alpha,q_\beta,q_\gamma,q_\delta\right\}$,
$\bar{J}=\left\{J_\alpha,J_\beta,J_\gamma,J_\delta\right\}$ and
$\overline{|m|}=\left\{|m_\alpha|,|m_\beta|,|m_\gamma|,|m_\delta|\right\}$.
The overall interaction coefficient can then be written as the product of Eqs.
\eqref{eq:U7} and \eqref{eq:U3}
\begin{multline}
  \label{eq:Uf}
  U_{\alpha,\beta,\gamma,\delta}=
                     \delta_{\|\bar{n}\|,2\mathbb{N}}
                     \delta_{m_\alpha+m_\beta,m_\gamma+m_\delta} 
              %\frac{\hbar^2a_s}{m l_x \pi^2 l_\perp^2}
\frac{{ \tilde a}_s }{4 l_x \pi^3 l_\perp^2}
                      \sum_{\bar{q}=\overline{|m|}}^{\bar{J}}
                          \sum_{\bar{s}}^{\bar{n}}
                             \frac{   \Lambda
                                           \left(
                                                 \bar{J},\bar{m},\bar{q}
                                           \right)
                                       \, \Xi \left(\bar{s}\right)}
                                  {\sqrt{2}^{\|\bar{s}\|+2\|\bar{q}\|}} \\
\times \Gamma \left ( \|\bar{q}\|+ \frac{3}{2} \right )\,
  \Gamma \left[
            \frac{ \left(\|\bar{n}\|-\|\bar{s}\|+1\right)| }{2}
         \right] \, .
\end{multline}
We are thus enabled to rewrite Hamiltonian 
\eqref{eq:ft-hamiltonian3} in terms of the calculated
coefficients obtaining
\begin{equation}
  \label{eq:ft-hamiltonian4}
  \hat{\mathsf{H}}=
                   \sum_{j}
                      \left[
                              \sum_{\alpha} 
                                   \lambda_{\alpha,\beta}
                                 \hat{n}^\dagger_\alpha+ 
                              \sum_{\alpha,\beta} 
                                    T_{\alpha,\beta} 
                                   \hat{c}^\dagger_\alpha
                                   \hat{c}_\beta
                       +\sum_{\alpha,\beta,\gamma,\delta} 
                                     U_{\alpha,\beta,\gamma,\delta}
                                   \hat{c}^\dagger_\alpha
                                   \hat{c}^\dagger_\beta
                                   \hat{c}_\delta
                                   \hat{c}_\gamma
  \right]\, .
\end{equation}
We will refer to \eqref{eq:ft-hamiltonian4} as the generalised Hubbard
Hamiltonian.  
For sake of simplicity, in Eq. \eqref{eq:ft-hamiltonian4}
we have not written down explicitly the selection rules
imposed by symmetry constraints (see below).

%%%%%%%%%%%%%%%%%%%%%%%%%%%%%%%%%%%%%%%%%%%%%%%%%%%%%%%%%%%%%%%%%%%%%%%%%%%%%%%
%%%%%%%%%%%%%%%%%%%%%%%%%%%%%%%%%%%%%%%%%%%%%%%%%%%%%%%%%%%%%%%%%%%%%%%%%%%%%%%
\subsection{Symmetry properties of the interaction term}
\label{sec:symm}
   
In addition to global symmetry properties, such as 
1) rotational symmetry along the
$x$-axis and 2) left-right symmetry, reflected by momentum $x$-component 
conservation and parity conservation for the 1D harmonic oscillators along the
$x$-axis, it is clear from equation \eqref{eq:Uf} that the coefficient
$U_{\alpha,\beta,\gamma,\delta}$ has some symmetry properties: 
a) $U$ does not depend on the sign of $m_\chi$ with
$\chi=\alpha,\beta,\gamma,\delta$, 
provided the conservation $m$ during the interaction 
($m_\alpha+m_\beta = m_\gamma+m_\delta$); 
b) $U$ possesses a permutational symmetry,
namely
\begin{equation}
    \label{eq:simmU}
    U_{\alpha,\beta,\gamma,\delta}=
      U_{\beta,\alpha,\gamma,\delta}
    =U_{\alpha,\beta,\delta,\gamma}=
      U_{\beta,\alpha,\delta,\gamma} \, .
\end{equation}

We would like to draw reader's attention to the two $\delta$ functions in
equation \eqref{eq:Uf} which make explicit the conservation laws that might
have been expected by simply considering the symmetry of the problem. The
first one represents parity conservation, while the second conservation of the
$x$ component of the angular momentum. In table \ref{tab:ut}, we give the
analytical value of $U$ for interaction between particles belonging to the
first three shells of the 2D radial harmonic oscillator and to the first level
for the axial harmonic oscillator. These symmetry constraints allow to class
the possible quantum numbers of the interacting particles according to the
value of the corresponding $U$. For example for
$$ 
\alpha=\{0,1,0,i,\sigma\}\, , \quad
\beta=\{0,0,0,i,\sigma'\}\, ,
$$
$$
\delta=\{0,0,0,i,\sigma\}\, , \quad
\gamma=\{0,0,0,i,\sigma'\}\, ,
$$ 
and 
$$ \alpha=\{0,0,0,i,\sigma\}\, , \quad
\beta=\{0,0,0,i,\sigma'\} \, , 
$$
$$
\delta=\{0,1,0,i,\sigma\}\, , \quad
\gamma=\{0,0,0,i,\sigma'\} \, ,
$$ 
we have the same value of $U$, henceforth the
class definition of table \ref{tab:ut}. We would like to point out two aspects
of this example. First of all it may be noticed that the angular momentum $J$
is not conserved throughout the interaction: this is a general feature of the
system considered, there is not global rotational symmetry but only in the
plane orthogonal to the 1D optical lattice. Moreover in this particular
interaction the value for the coefficient $U$ is negative, this appears to be
a rare (but not unique) situation. The implications of this condition will be
pointed out in section \ref{sec:SpCa}  
%
%%%%%%%%%%%%%%%%%%%%%%%%%%%%%%%%%%%%%%%%%%%%%%%%%%%%%%%%%%%%%%%%%%%%%%%%%%%%%%%%%%
\renewcommand {\arraystretch}{1.3}
%
%%%%%%%%%%%%%%%%%%%%%%%%%%%%%%%%%%%%%%%%%%%%%%%%%%%%%%%%%%%%%%%%%%%%%%%%%%%%%%%%%
\begin{table}[htbp]
  \centering
 \caption{Values of $\tilde{U}_{\alpha,\beta,\gamma,\delta}=
                     \left[\hbar^2a_s /(ml_x \pi^2 l_\perp^2)\right]^{-1}
                     U_{\alpha,\beta,\gamma,\delta}$ for
                    $\{n_\alpha,n_\beta,n_\gamma,n_\delta\}=\{0,0,0,0\}$,
                    $i_\alpha=i_\beta$, $\sigma$ and $\sigma'$ satisfy symmetry constraints.
The value $\overline{m_\chi}$ represents the equivalence class described in the text.}
\medskip
  \label{tab:ut} 
\begin{ruledtabular}
  \begin{tabular}{||c|c|c|c||c|c|c|c||c||}
   \hline  
    $J_\alpha$ & $J_\beta$ & $J_\gamma$ & $J_\delta$ & 
    $m_\alpha^*$ & $m_\beta^*$ & $m_\gamma^*$ & $m_\delta^*$ &
    $\tilde{U}_{\alpha,\beta,\gamma,\delta}$ \\ \hline 
    0 & 0 & 0 & 0 & 0 & 0 & 0 & 0 & $\frac{\pi}{2^4 }$ \\ \hline
    1/2 & 1/2 & 1/2 & 1/2 & 1/2 & 1/2 & 1/2 & 1/2 & $\frac{15\pi}{2^8}$ \\ \hline
    1 & 1 & 1 & 1 & 1 & 1 & 1 & 1 & $\frac{945\pi}{2^{14}}$ \\ \hline
    1 & 1/2 & 1/2 & 1 & 1 & 1/2 & 1/2 & 1 & $\frac{105\pi}{2^{11}}$ \\ \hline
    1 & 1 & 1 & 1 & 0 & 0 & 0 & 0 & $\frac{193\pi}{2^{12}}$ \\ \hline
    1/2 & 0 & 1/2 & 0 & 1/2 & 0 & 1/2 & 0 & $\frac{3\pi}{2^6}$ \\ \hline
    1 & 1 & 1 & 1 & 1 & 0 & 1 & 0 & $\frac{345\pi}{2^{13}}$ \\ \hline
    1 & 1/2 & 1/2 & 1 & 0 & 1/2 & 1/2 & 0 & $\frac{33\pi}{2^{10}}$ \\ \hline
    1 & 1/2 & 1/2 & 1 & 0 & 1/2 & -1/2 & 1 & $\frac{45\pi}{2^{10}\sqrt{2}}$ \\ \hline
    1 & 0 & 1 & 0 & 1 & 0 & 1 & 0 & $\frac{15\pi}{2^{9}}$ \\ \hline
    1 & 1 & 0 & 0 & 0 & 0 & 0 & 0 & $\frac{7\pi}{2^8}$ \\ \hline
    1 & 1 & 1 & 0 & 1 & -1 & 0 & 0 & $\frac{45\pi}{2^{10}}$ \\ \hline
    1 & 1/2 & 1/2 & 0 & 0 & 1/2 & 1/2 & 0 & $\frac{3\pi}{2^8}$ \\ \hline
    1 & 0 & 0 & 0 & 0 & 0 & 0 & 0 & $-\frac{\pi}{2^6}$ \\ 
\hline
\end{tabular}
\end{ruledtabular}
\end{table}
%%%%%%%%%%%%%%%%%%%%%%%%%%%%%%%%%%%%%%%%%%%%%%%%%%%%%%%%%%%%%%%%%%%%%%%%%%%%%%%%

\section{Special Cases}
    \label{sec:SpCa} 
    In this section we derive three model Hamiltonians for fermions
    in optical lattices. We consider, for both cases, only the
    lowest-state axial quantum number (i.e. $n_\alpha=n_\beta=0$).
    Hence $T_{\alpha,\beta}$ can be written as
  \begin{equation}
    \label{eq:TSp}
     T_{\alpha,\beta}=
                      \delta_{J_\alpha,J_\beta}\delta_{m_\alpha,m_\beta}
                      \delta_{\sigma_\alpha,\sigma_\beta}
                      T_{0,0,i_\alpha,i_\beta}
  \end{equation}
  
  As far as an ultracold gas is considered, it seems feasible to restrict our
  analysis to the first few levels above the ground state (i.e.
  $J_\alpha=0,1/2,\ldots$).  As a first example, we consider the case having
  $J_\alpha=0$ as the only radial level allowed and the fermionic gas is spin
  unpolarised. From Eq.  \eqref{eq:Uf}, along with the previous assumptions,
  we obtain \cite{Ruuska}
  \begin{equation}
    \label{eq:Hubb}
    \hat{H}= \sum_{i,\sigma} 
                 \mu_{i}
                \hat{n}_{i,\sigma}
         - T \sum_{i,\sigma}
              \left(
                     \hat{c}^\dagger_{i,\sigma}
                     \hat{c}_{i+1,\sigma}+
               \hat{c}^\dagger_{i+1,\sigma}
                       \hat{c}_{i,\sigma}\right)+ 
           U  \sum_{i,\sigma,\sigma'}
                   \hat{n}_{i,\sigma}
                   \hat{n}_{i,\sigma'}
  \end{equation}
  with $T=T_{0,0,i_\alpha,i_\alpha+1}$ which is easily recognised as the
  Hubbard Hamiltonian, whose role in the ultracold atoms physics has been
  pointed out elsewhere \cite{Jaksch, Hofstetter}.
%%%%%% 
  Note that in this example we have made the assumption that the tunnelling
  coefficient is significantly different from zero only for
  nearest-neighbouring sites. Nevertheless more involved situations may arise,
  suggesting interesting physical features, as it is shown in Appendix
  \ref{sec:AppA}.
  
  If we now consider a spin-polarised gas in a (radial) multi-level system we
  obtain \cite{Ruuska} 
  \begin{equation}
    \label{eq:SpPol}
    \hat{H}=
            \sum_{\bar{n},i}
               \mu_{\bar{n},i}\hat{n}_{\bar{n},i}-
                   T \sum_{\bar{n},i}
                 \left(\hat{c}^\dagger_{\bar{n},i}
                       \hat{c}_{\bar{n},i+1}+
                       \hat{c}^\dagger_{\bar{n},i+1}
                       \hat{c}_{\bar{n},i}\right)
  \end{equation}
  where the absence of the interaction term is related to the symmetry
  properties of the coefficient $U_{\alpha\beta\gamma\delta}$.
%%%%%%%%%% 
  The Hamiltonian \eqref{eq:SpPol} is readily diagonalised to yield
  \begin{equation}
    \label{eq:SpPolDiag}
    \hat{H}= \sum_{\bar{n}} \hat{H}_{\bar{n}} = 
     \sum_{\bar{n},k,\sigma} \left[ \mu_{\bar{n}} - 2T \cos(k) \right]
    \hat{n}_{\bar{n},k,\sigma} 
  \end{equation}
  with the same procedure followed in the strong coupling limit in the Hubbard
  Hamiltonian.

\subsection*{Rotational Hubbard Hamiltonian}
\label{sec:RMMHH}
  As the last example, we derive a third Hamiltonian that may give the
  reader a first insight on the increasing complexity if higher 
  single-particle levels are taken into account. Here we consider a situation
  where we allow $J_\alpha= 0, 1/2$ (but always $n_\alpha=0$, leading to
  $T_{ n_\alpha, i_\alpha, n_\beta, i_{\alpha+1}}=T$ as already pointed out)      
\begin{multline}
\label{eq:2-Level1}
  H_{2-level}=\sum_{i,\sigma} \sum_{a=-1}^1 
        \left[ 
             \lambda_{a} n_{i,a,\sigma}  - 
          T \left( 
                 c^\dagger_{i,a,\sigma}c_{i+1,a,\sigma}+  
                 c^\dagger_{i+1,a,\sigma} c_{i,a,\sigma}
           \right)
         \right] + \\ 
          \sum_{i} \sum_{a,b,c,d}
              U_{a,b,c,d}
              \sum_{\sigma,\sigma'} c^\dagger_{i,a,\sigma} c^\dagger_{i,b,\sigma'} 
                c_{i,c,\sigma'} c_{i,d,\sigma}    
\end{multline}
where $
U_{a,b,c,d}= 
U_{a,i_\alpha;b,i_\beta;
c,i_\gamma;d,i_\delta}
$. The label $a$ (as well as $b$, $c$, and $d$) has been introduced to represent
the triplet of harmonic-oscillator numbers $(n_\alpha, J_\alpha, m_\alpha)$, hence
$\alpha=\{a ,i_\alpha, \sigma_\alpha \}$. One should recall that originally
$\alpha = (n_\alpha, J_\alpha, m_\alpha , i_\alpha, \sigma_\alpha )$. 
Here, however, it is convenient to write in an explicit way
both spin indices $\sigma_\alpha$'s and site indices $i_\alpha$'s.

The triplet $a = (n_\alpha, J_\alpha, m_\alpha)$ is such that
the value $a=0$ corresponds to
$(0,0,0)$, $a=1 \to (0,1/2,+1/2)$ and $a=-1 \to
(0,1/2,-1/2)$. The axial quantum number $n_x$ has been ``frozen'' to 0 due to
the disk-shaped potential form (i.e. $\omega_x \gg \omega_\perp$) while the
radial quantum number $J$ has been limited to the values $\{0,1/2\}$ as a
first approximation beyond the $J =0$ (Hubbard Hamiltonian see \eqref{eq:Hubb}).
The present model thus enriches the dynamical scenario
by introducing modes that takes into account 
the simplest possible rotational processes
for fermions confined in a well.

The wealth of the scenario depicted in Eq. \eqref{eq:2-Level1} arises from the
level-dependence of the interaction coefficient 
$U_{a,b,c,d}$. In fact $U_{a,b,c,d}$,  as a function of the energy
levels may provide a useful tool to simplify Eq. \eqref{eq:2-Level1} hinting
the best strategy for both numerical and analytical analysis of this model. 
Two main aspects concerning these coefficients are worth repeating
here: a) the $m_\alpha$-conserving nature of the interaction, related to the
symmetry properties of the confining potential and of the interaction coefficient 
(see section \ref{sec:symm}), reduces the number of possible processes; b) the
symmetry properties of  $U_{\alpha,\beta,\gamma,\delta}$ (see Eq.
\eqref{eq:simmU}) allow the grouping of interaction terms, accordingly to what
has been done in table \ref{tab:ut}.

From a general point of view, in Hamiltonian \eqref{eq:2-Level1} the hopping
factor may be construed as a multichannel tunnelling coefficient, where the
radial quantum numbers identify the channel label, in the same spirit of
Hamiltonian \eqref{eq:SpPolDiag}. Incidentally, this is true if axial degrees
of freedom are ``frozen'' to $n_\alpha=n_\beta=0$, otherwise there is
tunnelling among levels with $n_\alpha \neq n_\beta$, for some $\alpha$ and
$\beta$. 

Hamiltonian \eqref{eq:2-Level1} can represent a situation where single traps
are loaded with a small number of atoms, as to fill the first two radial
levels of the local harmonic oscillator.  To experimentally obtain one of the
different simplified Hamiltonians -- like \eqref{eq:2-Level1}-- it is necessary
to have control of four experimental parameters: laser intensity, angle
between counterpropagating lasers, axial magnetic trapping frequency,
scattering length. With these parameters it is possible to gain full knowledge
of ``lattice constant'', interaction parameter, shape and depth of the 3D
harmonic traps.  The most critical point seems the few-atoms loading of the
trap but a technique involving a 3D anisotropic array --a sort of 2D array of
1D arrays-- might overcome the problem.

In this picture, the interaction coefficient $U$ can then be used as a source
of entanglement between different channels. Moreover the possibility of
experimental control of the scattering length, and thus of the interaction
term, via an applied magnetic field may provide an useful tool of external
manipulation of the state of the system in the rich scenario here depicted.

%%%%%%%%%%%%%%%%%%%%%%%%%%%%%%%%%%%%
%                                  %
%   new                            %
%                                  %
%%%%%%%%%%%%%%%%%%%%%%%%%%%%%%%%%%%% 

  To outline future paths of research, we will here sketch a way to set up 
  a mean-field procedure for the Rotational Hubbard Hamiltonian. The main 
  interest of this approach resides in the possibility  of a general
  discussion of some features of the model which have a direct
  experimental relevance. For example it is possible to state that, according
  to what is usually affirmed in the literature \cite{Lieb} , no BCS-like 
  ground-state is possible for repulsive interaction, while for an attractive
  two-body potential a paired ground state is possible. The flexibility of
  experimental techniques involved in the study of ultracold atom physics
  allows to envisage experimental conditions where these two different regimes
  are attained. For example exploiting a Feshbach resonance it is possible to
  drive the scattering length $a_s$ from positive to negative values leading 
  thus the system through a quantum phase transition.       
  
  The analytic procedure adopted hereafter deeply relies on the concept of 
  quasi-free state \cite{Lieb}. In our situation the following definition 
  of quasi-free state can be adopted: 
     \begin{itemize}
        \item[I)] all correlation functions can be computed from Wick's theorem;
        \item [II)] four fermionic expectation values over a quasi-free state
        have the form   
          \begin{eqnarray*} 
            &&<\phi|e_1 e_2 e_3 e_4|\phi>=
            <\phi|e_1e_2|\phi><\phi|e_3e_4|\phi>-\\  
            &&<\phi|e_1e_3|\phi><\phi|e_2e_4|\phi>+
            <\phi|e_1e_4|\phi><\phi|e_2e_3|\phi>\\
          \end{eqnarray*}
  
    \end{itemize}
  with $ e_i=c_i,\, c_i^\dagger$.
  In particular we would like point out how the three terms on the right
  hand-side will lead to the direct, the exchange and pairing energy term of a
  Hatree-Fock-Bogoliubov mean field Hamiltonian, which, for our RHH becomes
    \begin{eqnarray*}
      \label{eq:2HFB}
      \hat{H}_{2-level}^{HFB}&=&\hat{H}_0 + \sum_{\substack{i,a,b,c,d,\\ \sigma \neq \sigma'}}
                         U_{a,b,c,d}
                         \left[ 
                                \chi_{i,a,\sigma,c,\sigma}   \hat{c}^\dagger_{i,b,\sigma'} \hat{c}_{i,d,\sigma'} \right.\\
                            &+& \chi_{i,d,\sigma',b,\sigma'} \hat{c}^\dagger_{i,a,\sigma}  \hat{c}_{i,c,\sigma}  
                             -  \chi_{i,c,\sigma,b,\sigma'}  \hat{c}^\dagger_{i,a,\sigma}  \hat{c}_{i,d,\sigma'} \\
                            &-&  \chi_{i,d,\sigma',a,\sigma} \hat{c}^\dagger_{i,b,\sigma'} \hat{c}_{i,c,\sigma}  
                             + \xi^*_{i,b,\sigma',a,\sigma}  \hat{c}_{i,d,\sigma'}         \hat{c}_{i,c,\sigma} \\  
                    &+& \left. \xi_{i,c,\sigma,d,\sigma'}    \hat{c}^\dagger_{i,a,\sigma}  \hat{c}^\dagger_{i,b,\sigma'}
                         \right]
    \end{eqnarray*}
   with 
   \begin{eqnarray*}
      &&\hat{H}_0=\sum_{i,a,\sigma}
              \left[ 
                \lambda_a \hat{n}_{i,a,\sigma} 
                +
                T \left( 
                        \hat{c}^\dagger_{i+1,a,\sigma} \hat{c}_{i,a,\sigma}+ h.c.
                  \right)
              \right] \\
      &&\chi_{i,a,\sigma,b,\sigma'}=<\phi_{HFB}|\hat{c}^\dagger_{i,a,\sigma}
      \hat{c}_{i,b,\sigma'}|\phi_{HF}> \\
      &&\xi_{i,a,\sigma,b,\sigma'}=<\phi_{HFB}|\hat{c}_{i,a,\sigma}
      \hat{c}_{i,b,\sigma'}|\phi_{HFB}>
     \end{eqnarray*}
 
   The set of generators 
     $
       \left\{ \hat{c}^\dagger_\alpha \hat{c}_\beta-\frac{1}{2}\delta_{\alpha\beta}
       (1\leq \alpha \neq \beta\leq r),\right.$ 
     $ \left.\hat{c}_\alpha \hat{c}_\beta,
       \hat{c}^\dagger_\alpha \hat{c}^\dagger_\beta (1\leq \alpha \neq \beta \leq
       r) \right\}
     $
   obeys the following commutation relations
   \begin{eqnarray}
     \label{eq:sorComm}
     &&\left[\hat{c}^\dagger_i\hat{c}_j-\frac{1}{2}\delta_{ij},\hat{c}^\dagger_k\hat{c}_l-\frac{1}{2}\delta_{kl}\right]
     =\delta_{jk}(\hat{c}^\dagger_i\hat{c}_l-\frac{1}{2}\delta_{il})\nonumber\\
     &&\left[\hat{c}^\dagger_i\hat{c}_j-\frac{1}{2}\delta_{ij},
      \hat{c}^\dagger_k\hat{c}^\dagger_l \right]
     =\delta_{jk}\hat{c}^\dagger_i\hat{c}^\dagger_l-\delta_{jl}\hat{c}^\dagger_i\hat{c}^\dagger_k \nonumber\\
    &&\left[\hat{c}_i\hat{c}_j,\hat{c}^\dagger_k\hat{c}^\dagger_l\right]=\delta_{ik}
         (\hat{c}^\dagger_i\hat{c}_j-\frac{1}{2}\delta_{ij})+ \nonumber\\
         && \delta_{ij}(\hat{c}^\dagger_k\hat{c}_i
         -1/2\delta_{ki})-\delta_{li}(\hat{c}^\dagger_k\hat{c}_j-1/2\delta_{kj})\nonumber \\
         &&     \delta_{ki}(\hat{c}^\dagger_l\hat{c}_i-1/2\delta_{li})
    \end{eqnarray}  
 allowing to state that the dynamical algebra of this new Hamiltonian, 
 which is now quadratic in terms of $\hat{c}_i,
  \hat{c}_i^\dagger$  can be easily recognised to be
 $so(2r)$ \cite{Gilmore}.

  Having determined the dynamical algebra of the model Hamiltonian, enables
  us -- at least in principle- to find the ground state of the system with a
  straightforward procedure. As it will be clear form the subsequent
  discussion, the main difficulties arise  as the number the generators 
  of the $so(2r)$ algebra grows with $r(2r-1)$. For instance
  for the two-site, $J=0,1/2$ model, the Hamiltonian dynamical algebra 
  will have 276 generators.

  In spite of the technical difficulties (both analytical and numerical), it
  is appropriate to apply algebraic techniques to diagonalise
  $\hat{H}_{2-level}^{HFB}$. As stated before, this general approach will give some
  insight to the ground state properties of the system. 
  If we consider a unitary transformation $g \in SO(2r)$ we can write
  \begin{equation}
    \label{eq:diag1}
    \hat{H}_d=g\hat{H}_{HFB}g^{-1}
  \end{equation}
  where $\hat{H}_d$ is diagonal. As a direct consequence the ground state
  $|\phi_{HF}>$ of $\hat{H}_{2-level}^{HFB}$ can be written as
  \begin{equation}
    \label{eq:gsHFB}
     |\phi_{HF}>=g|0,0,0,\dots,0,0>=g|0>
  \end{equation}
  where $|0>$ can be defined as the Bogoliubov particle vacuum (ground state
  of $\hat{H}_d$). Following \cite{Gilmore}, $|0>$ represents a possible 
  choice for the extremal state for the $SO(2r)$ group with $U(r)$ as the
  corresponding maximum stability subgroup. Leading to
  \begin{equation}
    \label{eq:GsCohSt}
    g|0>=\Omega h |0> \Omega |0> e^{i\phi(h)}
  \end{equation}
  where:
  \begin{equation}
    \label{eq:CohStOp}
     \Omega=exp \sum_{1\leq \alpha \neq \beta \leq r}
             \left( \eta_{\alpha,\beta}\hat{c}^\dagger_\alpha
  \hat{c}^\dagger_\beta-H.c.\right) \in \frac{SO(2r)}{U(r)}. 
  \end{equation}
 The phase appearing in Eq.\eqref{eq:GsCohSt} has no relevance for our
 purposes, as we are interested in the evaluation of observable 
 expectation values.
 
 The problem mentioned above about the size of the dynamical algebra, appears here
 with all its implications. It is necessary to exponentiate the operator
 $ \sum_{1\leq \alpha \neq \beta \leq r}
             \left( \eta_{\alpha,\beta}\hat{c}^\dagger_\alpha
  \hat{c}^\dagger_\beta-H.c.\right)$ which is a $2r \times 2r$ matrix in the faithful
 matrix representation. 
 
 Nevertheless, for a repulsive two-body potential, the pairing term can be
 neglected \cite{Lieb}, thus the dynamical algebra of the system becomes $U(r)$.
 Following \cite{Gilmore}, we can express the Hamiltonian ground state as
 
 \begin{equation}
    |\phi_{HF}>=\exp \sum_{\substack{k+1\leq \alpha \leq r\\
                                  1 \leq j \leq k}}
             \left( \eta_{\alpha,\beta}\hat{c}^\dagger_\alpha
             \hat{c}_\beta-H.c.\right)|0>\label{eq:gsHF}
 \end{equation}
 where
 \begin{equation}
   \label{eq:ext}
   |0>=|\underbrace{1,1,\dots,1}_k,0,\dots,0>
 \end{equation}
 which is, in fact, the ground state of the non interacting Hamiltonian.
 It is worth noticing that this general procedure can be greatly simplified if
 further constraints, related to symmetries of the problem, are imposed onto
 the coefficients $\eta_{\alpha,\beta}$. For example, if we consider the
 two-site($A$,$B$), $J=0,1/2$ case, due to equation \eqref{eq:TSp} the matrix 
 $\overline{\eta}$ with elements $\eta_{\alpha,\beta}$ will have the form
 \begin{widetext}
 \begin{equation}
   \label{eq:formEta}
     \overline{\eta}=
     \begin{pmatrix}
       0&\eta_{1,2}&\eta_{1,3}&\eta_{1,4}&\eta_{1,5}&\eta_{1,6}&\eta_{1,7}&0&0&0&0&0\\
       -\eta_{1,2}&0&\eta_{2,3}&\eta_{2,4}&\eta_{2,5}&\eta_{2,6}&0&\eta_{2,8}&0&0&0&0\\
       -\eta_{1,3}&-\eta_{2,3}&0&\eta_{3,4}&\eta_{3,5}&\eta_{3,6}&0&0&\eta_{3,9}&0&0&0\\
       -\eta_{1,4}&-\eta_{2,4}&-\eta_{3,4}&0&\eta_{4,5}&\eta_{4,6}&0&0&0&\eta_{4,10}&0&0\\
       -\eta_{1,5}&-\eta_{2,5}&-\eta_{3,5}&-\eta_{4,5}&0&\eta_{5,6}&0&0&0&0&\eta_{5,11}&0\\
       -\eta_{1,6}&-\eta_{2,6}&-\eta_{3,6}&-\eta_{4,6}&-\eta_{5,6}&0&0&0&0&0&0&\eta_{6,12}\\
       -\eta_{1,7}&0&0&0&0&0&0&0&0&0&0&0\\
       0& -\eta_{2,8}&0&0&0&0&0&0&0&0&0&0\\
       0&0&-\eta_{3,9}&0&0&0&0&0&0&0&0&0\\
       0&0&0&-\eta_{4,10}&0&0&0&0&0&0&0&0\\
       0&0&0&0&-\eta_{5,11}&0&0&0&0&0&0&0\\
       0&0&0&0&0&-\eta_{6,12}&0&0&0&0&0&0\\
     \end{pmatrix}
 \end{equation}
 \end{widetext}
 thus impressively reducing the computational effort needed to evaluate the
 exponential in equation \eqref{eq:CohStOp}. In Eq. \eqref{eq:formEta} we have 
 have assumed the following convention
 \begin{eqnarray*}
   &&\{n_\alpha=0,J_\alpha=0,m_\alpha=0,i_\alpha=A,\sigma_\alpha=\uparrow \}\to 1 \\
   &&\{n_\alpha=0,J_\alpha=0,m_\alpha=0,i_\alpha=A,\sigma_\alpha=\downarrow \}\to 2 \\
   && \dots \\
   &&\{n_\alpha=0,J_\alpha=0,m_\alpha=0,i_\alpha=B,\sigma_\alpha=\uparrow \}\to 7 \\
   &&\{n_\alpha=0,J_\alpha=0,m_\alpha=0,i_\alpha=B,\sigma_\alpha=\downarrow\}\to 8 \\
   && \dots \\
 \end{eqnarray*}
 
%%%%%%%%%%%%%%%%%%%%%%%%%%%%%%%%%%%%
%                                  %
%  end new                         %
%                                  %
%%%%%%%%%%%%%%%%%%%%%%%%%%%%%%%%%%%% 

%%%%%%%%%%%%%%%%%%%%%%%%%%%%%%%%%%%%%%%%%%%%%%%%%%%%%%%%%%%%%%%%%%%%%%%%%%%%%%%%%
%%%%%%%%%%%%%%%%%%%%%%%%%%%%%%%%%%%%%%%%%%%%%%%%%%%%%%%%%%%%%%%%%%%%%%%%%%%%%%%%%
\section{Conclusions}
\label{sec:concl}
In this paper we have investigated the complex structure 
of fermion interactions for a fermion gas distributed in
a linear periodic array of potential wells. 
Based on the standard many-fermion quantum field theory 
endowed with a potential distribution mimicking a realistic
experimental setup, 
we have calculated analytically the hopping and interaction coefficients 
that describe the interactions of fermions within
a generalised multimode Hubbard Hamiltonian. Their dependence on
the external controllable parameters 
(such as laser intensity, magnetic trap frequency,
wavelength, and scattering length) has been determined.

Our analysis shows that,
except for two particularly simple cases (the gas of spin unpolarised fermions
and the gas of noninteracting spin polarised fermions), models with 
different degree 
of complexity can be derived depending on the interaction processes one decides
to account for or to neglect [consider, e. g., that, in principle, one might 
introduce an unlimited number of (local) rotational levels]. In this respect, 
our simplest nontrivial model (\ref{eq:2-Level1}), which is able to account 
for the (local) rotational activity of fermions, appears to be far more complex than
the Hubbard model or the spin-polarised noninteracting model derived 
in section \ref{sec:SpCa}. 
%
%resulting from the interplay of space, spin, and rotational quantum numbers
%
%In addition to investigate the zero-temperature properties
%of model (\ref{eq:2-Level1}), 

Therefore, the first objective of our future 
work is to perform a systematic study of model (\ref{eq:2-Level1}).
Based on the present analysis and exploiting the interaction-parameter 
scenario here depicted, the second objective
is to recognise the significant regimes characterising 
the confined fermion gas and to derive the relevant models 
from  Eq. \eqref{eq:ft-hamiltonian4}.

We would like to stress once again how the
analytical knowledge of the coefficients in principle allows us to tailor
Hamiltonians performing specific tasks.
  
An aspect that certainly deserves attention is the study of the
zero-temperature phase diagram of model (\ref{eq:2-Level1}) (and, more in
general, of sufficiently simple --and thus tractable-- models derived from the
gHH) and of the relevant phenomenology aimed at suggesting new possible
experiments.  To achieve a reliable description of these systems, several
established analytical and numerical approaches (see e.g. \cite{Rasetti,
  MontoPe}, and \cite{White, Batrouni,Freericks}, respectively) can be
implemented in analogy to what has been done for bosons
\cite{Batrouni,Freericks}. Moreover in the recent past several authors (see
e.g. \cite{Zanardi2, Viola}) have proposed to use entanglement measures as a
quantum phase transition identifier. We think that our model can represent a
good test-field for this new approach to quantum-phase transitions.

%%%%%%%%%%%%%%%%%%%%%%%%%%%%%%%%%%%%%%%%%%%%%%%%%%%%%%%%%%%%%%%%%%%%%%%%%%%%%
%%%%%%%%%%%%%%%%%%%%%%%%%%%%%%%%%%%%%%%%%%%%%%%%%%%%%%%%%%%%%%%%%%%%%%%%%%%%%%
  
\appendix
\section{Tunnelling coefficient calculation}
\label{sec:AppA}
In the following calculation we will fix $n_\beta \geq n_\alpha$, without loss
of generality, as it can be easily verified.

The integral in Eq. \eqref{eq:T5} can be decomposed in the sum of three terms 
\begin{equation}
  \label{eq:DecInt1}
  \Theta_1^{n_\alpha,n_\beta}+
  \Theta_2^{n_\alpha,n_\beta}+
  \Theta_3^{n_\alpha,n_\beta} = 
    \! \!  \int dy \,
             e^{-\frac{(y-\tau)^2}{2}}
             H_{n_\alpha}\left(y-\tau\right) 
                 f(y) 
             e^{-\frac{y^2}{2}}
             H_{n_\beta}\left(y\right)
%\\
\end{equation}

with  
$$ 
        f(y)=          \left[
                         \frac{1-\cos(2\Omega y)}{4\Omega^2} - \frac{y^2}{2}
                       \right]\, , 
$$
integral \eqref{eq:DecInt1} becomes 
\begin{equation}
\label{eq:DecInt2a}
                    \Theta_1^{n_\alpha,n_\beta} =
                      \int \! \frac{dy}{4\Omega^2}
                    e^{-\frac{y^2 +(y-\tau)^2 }{2} }
                       H_{n_\beta} (y-\tau ) H_{n_\alpha} ( y ) ,
\end{equation}
\begin{equation}
\label{eq:DecInt2b}
                      \Theta_2^{n_\alpha,n_\beta} =
                      \int \frac{dy}{4\Omega^2}
                     e^{-\frac{y^2 + (y-\tau)^2 }{2}} \, \cos(2\Omega y)  
\\
          \times H_{n_\alpha} \left(y\right) H_{n_\beta} \left(y-\tau\right) ,
\end{equation}
\begin{equation}
\label{eq:DecInt2c}
               \Theta_3^{n_\alpha,n_\beta}= - \frac{1}{2}
                     \int dy \, y^2 \,
                   e^{-\frac{y^2 + (y-\tau)^2}{2} }
%\nonumber
\\ 
           \times H_{n_\alpha} \left(y\right) H_{n_\beta} \left( y- \tau \right) ,
\end{equation}
The substitution $\zeta=y-\tau/2$ yields 
\begin{equation}
  \label{eq:Th1-1}
 \Theta_1^{n_\alpha,n_\beta} \! = 
                 C^{\tau}_{\Omega}
                   \int \! d\zeta 
                           e^{-\zeta^2} 
                             H_{n_\alpha} \! \left(\zeta+\frac{\tau}{2}\right)
                             H_{n_\beta} \! \left(\zeta-\frac{\tau}{2}\right)
\end{equation}
where $C^{\tau}_{\Omega}= e^{-{\tau^2}/{4} }/(4\Omega^2)$.
We then use the Hermite polynomial identity 
\begin{equation}
  \label{eq:Th1-2}
  H_n(x+y)=\sum_{k=0}^n {n \choose k}H_k(x)(2y)^{n-k}
\end{equation}
to obtain 
\begin{equation}
  \label{eq:Th1-3}
  \Theta_1^{n_\alpha,n_\beta} = 
                                \frac{e^{-\frac{\tau^2}{4}}}
                                     {4\Omega^2} 
                               \int d\zeta 
                                      e^{-\zeta^2} 
                                  \sum_{l,k=0}^{n_\alpha ,n_\beta }
                                        {n_\alpha \choose l}
                                        {n_\beta \choose k}         
                                          \tau^{n_\alpha +n_\beta
                                          -(l+k)}(-1)^{n_\beta -k}H_k(\zeta)H_l(\zeta)
\end{equation}
With the orthogonality of Hermite polynomials 
\begin{equation}
  \label{eq:Th1-4}
  \int_{-\infty}^\infty dx H_n (x)H_m (x) e^{-x^2} = \delta_{n,m} 2^n  n! \sqrt{\pi}
\end{equation}
we are able to perform the $\zeta$ integration 
\begin{equation}
  \label{eq:Th1-5}
  \Theta_1^{n_\alpha,n_\beta} = 
                               \frac{\sqrt{\pi}}
                                    {4\Omega^2}
                                    e^{-\frac{\tau^2}{4}} (-1)^{n_\beta}
                                 \sum_{l=0}^{n_\alpha }
                                    {n_\alpha  \choose l}
                                    {n_\beta  \choose l} 
                              \tau^{n_\beta +n_\alpha -2l} (-2)^l l! \, .
\end{equation}
It is worth noting that the summation extends 
to $n_\alpha$, that is ${a \choose b}=0$ if $a<b$.
From \cite{Wunsche} it can be verified that the last summation is related to
generalised Laguerre polynomials, giving 
\begin{equation}
  \label{eq:Th1-6}
  \Theta_1^{n_\alpha,n_\beta} = 
        \frac{\sqrt{\pi} n_\alpha ! 2^{n_\alpha} }{4\Omega^2 (-\tau)^{n_\alpha -n_\beta }}
                e^{-\frac{\tau^2}{4}}  
                        L_{n_\alpha} ^{n_\beta -n_\alpha}
                       \left(\frac{\tau^2}{2}\right) \, .
\end{equation}

We move now to the calculation of $\Theta_2^{n_\alpha,n_\beta}$which is given
by 
\begin{equation}
  \label{eq:Th2-1}
  \Theta_2^{n_\alpha,n_\beta}= 
                        -\frac{1}{4\Omega^2} 
                   \int dy \exp\left[-\frac{y^2}{2}\right]
                           H_{n_\alpha}\left(y\right) 
                       \exp\left[-\frac{(y-\tau)^2}{2}\right]
                           H_{n_\beta}\left(y-\tau\right)
                     \cos(\Omega y) \, .
\end{equation}
Eq. \eqref{eq:Th2-1}, with the substitution $\zeta=y-\tau/2$, can be written
as 
\begin{equation}
  \label{eq:Th2-2}
  \Theta_2^{n_\alpha,n_\beta}= 
                               -\frac{1}
                                     {4\Omega^2}
                                \exp\left(-\tau^2/4\right)
                                   \int d \zeta  e^{-\zeta^2} 
                                           H_{n_\alpha} 
                                          \left(\zeta + \frac{\tau}{2}\right) 
                                                 H_{n_\beta} 
                                    \left(
                                           \zeta + \frac{\tau}{2}
                                    \right) 
                                    \cos\left(\Omega \zeta \right) \, .
\end{equation}
Again, using Eq.\eqref{eq:Th1-2} gives
\begin{multline}
  \label{eq:Th2-3}
  \Theta_2^{n_\alpha,n_\beta}=
                              -\frac{e^{-\tau^2/4}}
                                    {4\Omega^2}
                              \sum_{l,k=0}^{n_\alpha ,n_\beta }
                                    {n_\alpha \choose l}
                                    {n_\beta \choose k}    
                                       \tau^{n_\beta +n_\alpha -(l+k)} \\
                           \times (-1)^{n_\beta -k} 
                  \int d\zeta e^{-\zeta^2} H_k(\zeta)H_l(\zeta) \cos\left(2\Omega \zeta \right)
\end{multline}
The integral in equation \eqref{eq:Th2-3} can be interpreted as the real
Fourier transform of the function 
\begin{equation*}
e^{-\zeta^2} H_k(\zeta)H_l(\zeta)\, .
\end{equation*}
Recalling that
\begin{equation*}
\mathcal{F}\left[f(x)g(x)\right]=\mathcal{F}
\left[ f(x) \right]*\mathcal{F}\left[g(x)\right] \, ,
\end{equation*}
where $\mathcal{F}\left[\cdot\right]$ indicates Fourier transform and $*$
convolution product, we obtain 
\begin{multline}
  \label{eq:Th2-4}
  \Theta_2^{n_\alpha,n_\beta}= 
-   \frac{e^{-\frac{\tau^2}{4 } }}{4\Omega^2}
                          \sum_{l,k=0}^{n_\alpha ,n_\beta }
                                {n_\alpha \choose l}
                                {n_\beta \choose k}  
                                \tau^{n_\beta+n_\alpha -l-k}  \\
                          \times (-1)^{n_\beta-k} \mathbb{R}e 
                                 \left[ \mathcal{F}
                                    \bigl [
                                           e^{-\frac{\zeta^2}{2} }
                                           H_k(\zeta)
                                    \bigr ] \!
                                        * \! \mathcal{F}
                                    \bigl [
                                           e^{-\frac{\zeta^2}{2} } H_l(\zeta)
                                    \bigr ]
                                 \right]
\end{multline}
giving
\begin{multline}
  \label{eq:Th2-5}
  \Theta_2^{n_\alpha,n_\beta}= 
           -\frac{e^{-\tau^2/4}}{4\Omega^2} 
                  \sum_{l,k=0}^{n_\alpha ,n_\beta }
                    {n_\alpha \choose l}
                    {n_\beta \choose k}
                      \tau^{n_\beta +n_\alpha -(l+k)} (-1)^{n_\beta -k+l} \\
     \times \mathbb{R}e \left[ 
                         i^{k+l} \int d\epsilon 
                              e^{-\frac{\epsilon^2 +(\epsilon-2\Omega)^2}{2}}
                                          H_k(\epsilon)
                                          H_l (\epsilon-2\Omega )
                       \right]\, .
\end{multline}
The integral on the left-hand side of equation \eqref{eq:Th2-4} can be solved
following the same procedure used for $\Theta_1^{n_\alpha,n_\beta}$
\begin{equation}
  \label{eq:eq:Th2-5-2}
  \int d\epsilon e^{-\epsilon^2/2}
                       H_k(\epsilon)
                         e^{-\left(\epsilon-2\Omega\right)^2/2}
                       H_l\left(\epsilon-2\Omega\right)= 
               (-1)^{k-l}\sqrt{\pi} e^{-\Omega}\left(2\Omega\right)^{k-l} l! 2^l
                                     L_l^{k-l}\left(4\Omega^2\right) 
\end{equation}
giving
\begin{multline}
  \label{eq:eq:Th2-6}
  \Theta_2^{n_\alpha,n_\beta}= -\frac{ e^{-\frac{\tau^2}{4} } }
                                     {4\Omega^2}
                                \sum_{l,k=0}^{n_\alpha ,n_\beta }
                                        {n_\alpha \choose l}
                                        {n_\beta  \choose k}
\tau^{n_\beta +n_\alpha -(l+k)} (-1)^{n_\beta}
                            \\ \times
                          2^l l!   
                                 \, \mathbb{R}e 
                                 \left[ i^{k+l} \right]
                                \frac{ \sqrt{\pi} }{ ( 2\Omega )^{l-k} }
L^{k-l}_{l} \left(2\Omega^2\right) e^{-\Omega^2} \, .
\end{multline}
The calculation of $\Theta_3^{n_\alpha,n_\beta}$ is quite straightforward.
Applying twice the identity
\begin{equation}
  \label{eq:Th3-1}
  xH_n(x)=\frac{1}{2}H_{n+1}(x)+nH_{n-1}(x)
\end{equation}
we can write $\Theta_3^{n_\alpha,n_\beta}$ as
\begin{multline}
  \label{eq:Th3-2}
  \Theta_3^{n_\alpha,n_\beta} = - \frac{e^{-\frac{\tau^2}{4}}}{2}
                       \int d\zeta
                           e^{-\zeta^2} 
                               \left [
                                     \frac{1}{4}
                                     H_{n_\alpha +2}
                                        \left( \zeta+\frac{\tau}{2} \right)+ 
                                        \frac{2 n_\alpha +1}{2} \,
                                     H_{n_\alpha}
                                        \left( \zeta+\frac{\tau}{2} \right)+ 
                                     n_\alpha (n_\alpha -1)  \right.  \\                                      
\left. \times H_{n_\alpha -2}   \left(\zeta+\frac{\tau}{2} \right)
                               \right ]
                           H_{n_\beta} \left(\zeta-\frac{\tau}{2}\right) \, .
\end{multline}
With the same procedure used for $\Theta_1^{n_\alpha,n_\beta}$,
$\Theta_3^{n_\alpha,n_\beta}$ is given by
\begin{multline}
 \label{eq:Th3-3}
 \Theta_3^{n_\alpha,n_\beta} = \frac{(-1)^{n_\beta -n_\alpha +1}}{2}
                                 \sqrt{\pi}
                                 2^n_\alpha 
                                 n_\alpha ! \,
                                 \tau^{ n_\beta -n_\alpha }
                                 e^{ -\frac{\tau^2}{4} }  
                               \left [
                       \frac{ (n_\alpha +1)(n_\alpha +2) }{\tau^{2}}
                                      L_{n_\alpha +2}^{ n_\beta- n_\alpha -2}
                                 ( {\tau^2}/{2} ) \right. +  
\\
+ \left.
\frac{ \tau^2 }{4} L_{n_\alpha-2}^{n_\beta -n_\alpha +2} \left ( \frac{\tau^2}{2} \right )
                 +  \frac{2n_\alpha +{1}}{2} 
               L_{n_\alpha}^{ n_\beta - n_\alpha } \left( \frac{\tau^2}{2} \right ) 
\right ] \, . 
\end{multline}
Hence $T_{\alpha,\beta}$ becomes
\begin{equation}
  \label{eq:T-f}
  T_{\alpha,\beta}= -\frac{\hbar
                          \omega_x \delta_{J_\alpha,J_\beta}
                         \delta_{m_\alpha , m_\beta} 
                         \delta_{\sigma_\alpha , \sigma_\beta}}
                  {\sqrt{2^{n_\alpha+n_\beta+2} \, n_\alpha! \, n_\beta! \, \pi}}
                          \left[ 
                                 \Theta_1^{n_\alpha,n_\beta}+
                                 \Theta_2^{n_\alpha,n_\beta}+
                                 \Theta_3^{n_\alpha,n_\beta}
                          \right] 
\end{equation}
with $\Theta_1^{n_\alpha,n_\beta}$, $\Theta_2^{n_\alpha,n_\beta}$,
and $\Theta_3^{n_\alpha,n_\beta}$ given by formulas 
(\ref{eq:Th1-6}), (\ref{eq:eq:Th2-6}), and (\ref{eq:Th3-3}),
respectively.

We have here the plot of $T_{n_\alpha,n_\beta,i_\alpha,i_{\alpha}+1}$ as a
function of the difference $i_\alpha-i_\beta$ for values of $n_\alpha$ and
$n_\beta$ ranging from 0 to 2.

\begin{figure}[htbp]
    \epsfig{figure=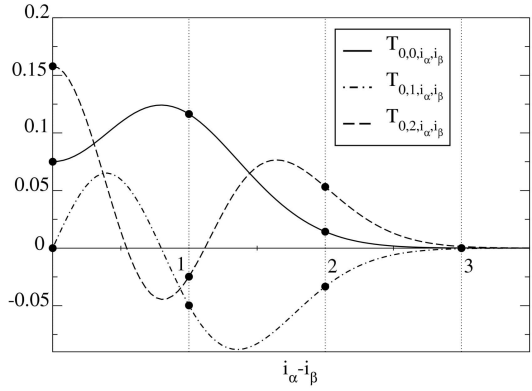,angle=270,width=0.5\textwidth}
      \caption{Plot of $T_{n_\alpha,n_\beta}$ from
        $T_{0,0,i_\alpha,i_\beta}$ to $T_{0,2,i_\alpha,i_\beta}$.  The solid
        line represents the case of a ground-state tunnelling. In this case the
        hopping parameter $T_{0,0,i_\alpha,i_\beta}$ is always
        positive. However, inter-level tunnelling already shows sign changes.}
      \label{t0xf} 
 \end{figure}

\begin{figure}[htbp]
     \centering
       \epsfig{figure=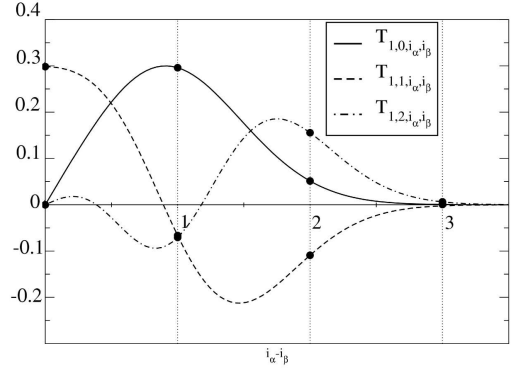,angle=270,width=0.5\textwidth}
        \caption{Plot of $T_{n_\alpha,n_\beta}$ from
                 $T_{1,0,i_\alpha,i_\beta}$ to $T_{1,2,i_\alpha,i_\beta}$. In
                 this situation the intra-level tunnelling term
                 $T_{1,1,i_\alpha,i_\beta}$ is always negative, but if
                 a different external parameter choice is considered, the
                 sign change can be placed between 1 and 2.} 
       \label{t1xf} 
 \end{figure}

\begin{figure}[htbp]
      \centering
       \epsfig{figure=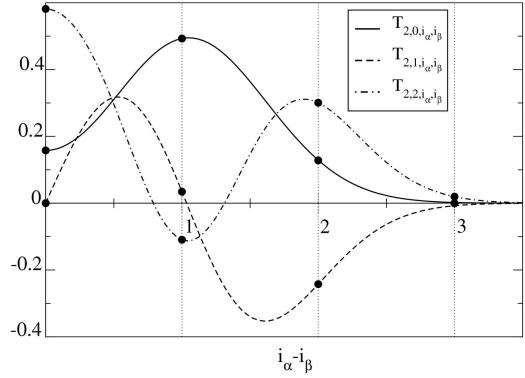,angle=270,width=0.5\textwidth}
         \caption{Plot of $T_{n_\alpha,n_\beta}$ from
           $T_{2,0,i_\alpha,i_\beta}$ to $T_{2,2,i_\alpha,i_\beta}$. In this
           case the sign change for the intra-level tunnelling term occurs for
           the specific parameter choice performed here, but it can be
           removed by a different choice of the external parameters.}
       \label{t2xf}      
 \end{figure}

The long distance exponential decay is common to all tunnelling coefficients,
regardless of the energy level. On the other hand its detailed shape has deep
relevance for nearest-neighbours and next-to-nearest-neighbours (i.e. there
may be sign changes passing from $T_{n,m,i,i+1}$ and $T_{n,m,i,i+2}$) as shown
in figures (\ref{t0xf}-\ref{t2xf}). Another interesting feature is that an
extra-term, due to ``on site'' tunnelling coefficients, bust be added to the
harmonic-oscillator energy term.   

%%%%%%%%%%%%%%%%%%%%%%%%%%%%%%%%%%%%%%%%%%%%%%%%%%%%%%%%%%%%%%%%%%%%%%%%%%%%%%%%%%%%%
%%%%%%%%%%%%%%%%%%%%%%%%%%%%%%%%%%%%%%%%%%%%%%%%%%%%%%%%%%%%%%%%%%%%%%%%%%%%%%%%%%%%%

\section{Interaction coefficients}
\label{sec:AppB}
We provide here the detailed calculation for the interaction term matrix
elements.
To solve integral \eqref{eq:U2}
\begin{equation*}
     U_x=\frac{1}{\pi l_{x}}
     \sqrt{\frac{2^{-(n_\alpha+n_\beta+n_\gamma+n_\delta)}}
              {n_\alpha!n_\beta!n_\gamma!n_\delta!}} 
       \times \int dx 
                H_{n_\alpha}(x)
                H_{n_\beta}(x) 
                H_{n_\gamma}(x)
                H_{n_\delta}(x)
                e^{-2x^2}
\end{equation*}
we exploit again Eq. \eqref{eq:Th1-2}, obtaining
\begin{multline}
  \label{eq:UB1}
  U_x=\frac{1}{\pi l_{x}}
     \sqrt{\frac{2^{-(n_\alpha+n_\beta+n_\gamma+n_\delta)}}
              {n_\alpha!n_\beta!n_\gamma!n_\delta!}}
            \sum
               {n_\alpha \choose i_\alpha} 
               {n_\beta  \choose i_\beta}
               {n_\gamma \choose i_\gamma}
               {n_\delta \choose i_\delta}
               H_{i_\alpha}(0)
               H_{i_\beta}(0)
               H_{i_\gamma}(0)
               H_{i_\delta}(0)
               \\ \times
               \int d\zeta 
               (2\zeta)^{n_\alpha +n_\beta+n_\gamma+n_\delta
               -(i_\alpha+i_\beta+i_\gamma+i_\delta)}e^{-2\zeta^2} \, .
\end{multline}
In previous equation the summation must be intended over four independent of
$s_\theta=0\dots n_\theta$ with $\theta=\alpha,\beta,\gamma,\delta$. 
With the substitution
\begin{equation}
  \label{eq:UB2}
     \int d\zeta (2\zeta)^\alpha 
e^{-2\zeta^2}=\delta_{\alpha,2\mathbb{N}}\sqrt{2}^{\alpha-3}
     \Gamma\left[(\alpha+1)/2\right]
\end{equation}
($\delta_{\alpha,2\mathbb{N}}$ indicates that $\alpha$ must be an even number)    
Eq. \eqref{eq:UB1} becomes
\begin{equation}
  \label{eq:UB3}
 U_x= \frac{1}{\pi l_x}\sum_{\bar{s}}^{\bar{n}}
   \frac{\Xi ({ \bar s} ) }{ \sqrt{2}^{\| \bar{s} \|+3}} 
   \Gamma\left[ \frac{ \|\bar{n}\|-\|\bar{s}\|+1 }{2} \right ]
\delta_{\|n\|,2\mathbb{N}}
\end{equation}
with
\begin{equation}
  \label{eq:UBfex}
  \bar{a}=\{a_\alpha,a_\beta,a_\gamma,a_\delta\} \mbox{\quad  1-norm: \quad}
  \|\bar{a}\|=\sum_\theta a_\theta
\end{equation}
and
\begin{equation}
  \label{eq:Xi}
  \Xi\left(\bar{s}\right)= \prod_\theta \frac{1}{\sqrt{n_\theta!}}{n_\theta \choose s_\theta}
  H_{s_\theta}(0) 
\end{equation}
where $ \Xi\left(\bar{s}\right)=0$ for odd $i_\theta$.
The $\delta$ function in 
Eq. \eqref{eq:UB3} should be written as 
$\delta_{(\|\bar{n}\|-\|\bar{s}\|,2\mathbb{N}}$. However the condition
$\|i_\theta\|=$even already implies $\|\bar{s}\|=$even. We are then allowed to
write in Eq. \eqref{eq:UB3}:
$$
 \delta_{\|\bar{n}\|-\|\bar{s}\|,2\mathbb{N}}=\delta_{\|\bar{n}\|,2\mathbb{N}} 
\, .
$$

We solve now the radial part of the interaction term integral written 
in Eq. \eqref{eq:U5}
\begin{equation*}
\! 
U_\rho=\int \! \! \int \,
\frac{d^2\eta}{\pi} \mathcal{L}^*_{J_\alpha,m_\alpha}(\eta)
                          \mathcal{L}^*_{J_\gamma,m_\gamma}(\eta)
                          \mathcal{L}_{J_\beta,m_\beta}(\eta)
                          \mathcal{L}_{J_\delta,m_\delta}(\eta) \, ,
\end{equation*}
where $\eta = (\rho,\phi)$ and $d^2\eta = \rho d\rho d\phi$.
Following \cite{Wunsche}, we express $\mathcal{L}_{J_\alpha,m_\alpha}(\rho,\phi)$
in terms of a finite sum:
\begin{multline}
  \label{eq:lSum}
  \mathcal{L}_{J_\alpha,m_\alpha}(\rho,\phi)
                           =e^{2i m_\alpha \phi}
                        e^{ -\frac{\rho^2}{l_\perp^2} }
{ \sqrt{(J_\alpha+m_\alpha)! (J_\alpha-m_\alpha)!} }
\\
                     \times \frac{ 1}{\pi}  \sum_{q_\alpha=|m_\alpha|}^{J_\alpha}
            \frac{(-1)^{J_\alpha-q_\alpha}
                        \left( \frac{\rho}{l_\perp}   \right )^{2q_\alpha}}
                 {(J_\alpha-q_\alpha)!
                         (q_\alpha+m_\alpha)!
                             \left(q_\alpha-m_\alpha\right)!}
\end{multline}
Hence the radial part of $\mathcal{L}_{J_\alpha,m_\alpha}(\tilde{\rho},\phi)$ can be written as
\begin{equation}
  \label{eq:defR}
  R_{J_\alpha,m_\alpha}(\tilde{\rho})
  =\frac{e^{-\left(\tilde{\rho}/l_\perp\right)^2/2}}
        {l_\perp}
        \sum_{q_\alpha=|m_\alpha|}^{J_\alpha} \Lambda_\alpha
           \left(\tilde{\rho}/l_\perp\right)^{2q_\alpha}
\end{equation}
where
\begin{equation}
  \label{eq:deflamb}
  \Lambda_\alpha = \frac{(-1)^{J_\alpha-q_\alpha}\sqrt{(J_\alpha+m_\alpha)!
                     (J_\alpha-m_\alpha)!}}
                    {(J_\alpha-q_\alpha)!
                     (q_\alpha+m_\alpha)!
                    \left(q_\alpha-m_\alpha\right)!} \, .
\end{equation}
Substituting Eq. \eqref{eq:defR} into Eq. \eqref{eq:U5} we have
\begin{multline}
  \label{eq:U2B2}
  U_\rho=\frac{2\delta_{m_\alpha+m_\gamma,m_\beta+m_\delta}}
              {\pi l_\perp^4}
           \sum_{q_\alpha=|m_\alpha|}^{J_\alpha} 
               \sum_{q_\beta=|m_\beta|}^{J_\beta}
                  \sum_{q_\gamma=|m_\gamma|}^{J_\gamma} 
                     \sum_{q_\delta=|m_\delta|}^{J_\delta} 
                \\ \times 
                   \Lambda_\alpha 
                   \Lambda_\beta 
                   \Lambda_\gamma 
                   \Lambda_\delta
             \int_0^\infty d\rho \frac{\rho}{\pi l_\perp^2} 
                         \left(\frac{\rho}{l_\perp}\right)^{2\|\bar{q}\|} 
                   e^{-2 { \tilde{\rho} }^2 }
\end{multline}
that, with the same notation of Eq. \eqref{eq:UB3}, becomes
\begin{equation}
  \label{eq:U2Bf}
 U_\rho= \frac{\delta_{m_\alpha+m_\gamma,m_\beta+m_\delta}}
              {\pi l_\perp^2}
         \sum_{\bar{q}=\overline{|m|}}^{\bar{J}}
           \Lambda\left(\bar{J},\bar{m},\bar{q}\right)  
           \frac{\Gamma \left( \|\bar{q}\| +\frac{3}{2} 
\right ) }{2^{\|\bar{q}\|+3/2}}
\end{equation}
with
\begin{equation}
  \label{eq:U2Badd}
  \Lambda\left(\bar{J},\bar{m},\bar{q}\right)=
       \prod_{\theta=\alpha,\beta,\gamma,\delta}
         \Lambda_\theta \, .
\end{equation}

\end{document}